\pdfoutput=1
\documentclass[11pt]{iopart}

\usepackage{amsfonts}
\usepackage{amssymb}
\usepackage{graphicx}
\usepackage{epstopdf}
\usepackage{verbatim}
\usepackage{color}
\usepackage{multirow}
\usepackage{mathcomp}
\usepackage{array}
\usepackage{bm}
\usepackage{wasysym}
\usepackage{setstack}
\usepackage{comment}
\usepackage[utf8]{inputenc}
\usepackage{csquotes}
\usepackage{harvard}
\usepackage{lineno}
\usepackage[dvipsnames]{xcolor}
\usepackage[normalem]{ulem}

\usepackage[thickspace,amssymb]{SIunits}

\newcommand{\he}{$^{\mathrm{4}}$He}
\newcommand{\car}{$^{\mathrm{12}}$C}
\newcommand{\oxi}{$^{\mathrm{16}}$O}
\newcommand{\betaplus}{$\beta^{\mathrm{+}}$}

\usepackage{amsfonts}
\usepackage{amssymb}
\usepackage{graphicx}
\usepackage{harvard}
\usepackage{epstopdf}
\usepackage{verbatim}
\usepackage{color}
\usepackage{xcolor}
\usepackage{subfigure}
\usepackage{wrapfig}
\usepackage{wasysym}
\usepackage{setstack}
\usepackage[thickspace,amssymb]{SIunits}
\usepackage{multirow}

\begin{document}

\title[Charged secondaries produced by $^{4}$He and $^{12}$C ion beams in a PMMA target]{Secondary radiation measurements for particle therapy applications: \newline Charged secondaries produced by $^{4}$He and $^{12}$C ion beams in a PMMA target at large angle}

\author{ A.~Rucinski$^{a,b,*}$, E.~De Lucia$^{c,\ddag}$, G.~Battistoni$^{d}$, F.~Collamati$^{c}$, R.~Faccini$^{a,e}$, P.~M.~Frallicciardi$^{f}$, C.~Mancini-Terracciano$^{a,e}$, M.~Marafini$^{a,g}$, I.~Mattei$^{d}$, S.~Muraro$^{d}$, R.~Paramatti$^{a,e}$, L.~Piersanti$^{a,b}$, D.~Pinci$^{a}$, A.~Russomando$^{a,e}$, A.~Sarti$^{b,c,g}$, A.~Sciubba$^{a,b,g}$, E.~Solfaroli~Camillocci$^{a,e}$, M.~Toppi$^c$, G.~Traini$^{a,e}$, C.~Voena$^{a}$, V.~Patera$^{a,b,g}$} 

\address{$^a$ INFN - Sezione di Roma, Italy}
\address{$^b$ Dipartimento di Scienze di Base e Applicate per Ingegneria, Sapienza Universit\`a di Roma, Roma, Italy}
\address{$^c$ Laboratori Nazionali di Frascati dell'INFN, Frascati, Italy} 
\address{$^d$ INFN - Sezione di Milano, Italy}
\address{$^e$ Dipartimento di Fisica, Sapienza Universit\`a di Roma, Roma, Italy}
\address{$^f$ Istituto di ricerche cliniche Ecomedica, Empoli, Italy}
\address{$^g$ Museo Storico della Fisica e Centro Studi e Ricerche ``E.~Fermi'', Roma, Italy}

\eads{(corresponding authors) *\mailto{antoni.rucinski@gmail.com}, $\ddag$\mailto{erika.delucia@lnf.infn.it}}
 
 
\begin{abstract}
Proton and carbon ion beams are used in the clinical practice for external radiotherapy treatments achieving, for selected indications, promising and superior clinical results with respect to X-ray based radiotherapy. Other ions, like \he are recently being considered as projectiles in particle therapy centres. \he~ions might represent a good compromise between the linear energy transfer and the radiobiological effectiveness of \car~ion and proton beams allowing improved tumour control probability and minimizing normal tissue complication probability. Proton, \he~and \car~ion beams allow to achieve sharp dose gradients on the boundary of the target volume. At the same time, the accurate dose delivery is more sensitive to the patient positioning and to anatomical variations with respect to photon therapy. This requires beam range and/or dose release measurement during the patient irradiation and therefore the development of dedicated monitoring techniques.

Measurements performed with the purpose of characterizing the charged secondary radiation for dose release monitoring in particle therapy are reported. Charged secondary yields, energy spectra and emission profiles produced in poly-methyl methacrylate (PMMA) target by \he~and \car~beams of different therapeutic energies were measured at 60\degree ~and 90\degree ~with respect to the primary beam direction. The secondary yields of protons produced along the primary beam path in PMMA target were obtained. The energy spectra of charged secondaries were obtained from time-of-flight information, whereas the emission profiles were reconstructed exploiting
tracking detector information. The measured charged secondary yields and emission profiles are in agreement with the results reported in literature and confirm the feasibility of ion beam therapy range monitoring using \car ~ion beam. The feasibility of range monitoring using charged secondary particles is also suggested for \he ~ion beam. 

\end{abstract}



\section*{Introduction}
\label{sec:intro}

The results of clinical studies support the application of proton and \car~ion beams for cancer treatment~\cite{Allen2012,Loeffler2013,Kamada2015}. In order to fully exploit the clinical advantages of Particle Therapy (PT) the research in medical physics focuses on increasing the benefits of ion beam therapy treatments. Recent considerations on advances in PT include the application of \he~ ion beams for more efficient treatment and increased life expectancy of pediatric patients. \he~ ion beams potentially exhibit properties which are a compromise between properties of protons and Carbon ions, particularly exhibit increased radiobiological effectiveness and suffer less lateral multiple scattering with respect to protons having a lower beam fragmentation with respect to carbon ions ~\cite{Kaplan1994,Castro1997,Tommasino2015a,Mairani2016,Mairani2016a,Kramer2016}.

Ions deposit the maximum dose at the end of their range in tissue, the Bragg Peak (BP), contrarily to photons that deposit their maximum dose close to the patient surface. The superposition of several Bragg curves creates the so-called Spread-Out BP (SOBP) covering the target volume with a homogeneous dose distribution achieving sharp dose gradients between the target region and the surrounding healthy tissue. Therefore the dose distributions obtained with ion beams are more conformal to the target volume with respect to those obtained with X-rays, due to the dose deposition characteristic and to the usage of active beam delivery method (\textit{i.e.} active beam scanning)~\cite{Haberer1993}. Also the increased radiobiological effectiveness of light ions that are heavier than protons makes PT favorable to treat radioresistant tumours~\cite{Tommasino2015,Paganetti2014}. 

On the other hand the scanned ion beam therapy is particularly sensitive to patient positioning and anatomical variations. Such variations may cause the BP position to be displaced during the treatment delivery with respect to the BP position predicted in the treatment plan, generating at the distal end of the SOBP what is commonly called the beam range uncertainty~\cite{Knopf2013}. 

In order to fully exploit the advantages of ion beams in the clinical practice, the development of novel techniques to verify and/or monitor the beam range in the patient during the therapy is demanded. In literature several monitoring strategies based on the measurement of secondary radiation exiting the patient were proposed, \textit{e.g.}, prompt gamma~\cite{Agodi2012a,Mattei2015,Mattei2016,Roellinghoff2014,Testa2014}, charged secondaries~\cite{Agodi2012b,Henriquet2012,Piersanti2014} and \betaplus~coincidence photon~\cite{Parodi2000,Agodi2012c,Kraan2014,Sportelli2014,Parodi2015}. Until now, none of these solutions were recognized to be clearly superior and/or universal. 

In this paper measurements of charged secondary particle
production induced by \he ~and \car ~beams at therapeutic energies impinging on a tissue-equivalent target made of polymethyl methacrylate (PMMA) are reported. The accurate measurement of the charged secondary particle yield is crucial to design a monitoring detector and optimize its positioning with respect to the primary beam direction. A precise knowledge of the number of secondary particles produced per primary ion is crucial also to achieve the required resolution on the emission profile reconstruction for PT dose monitoring. In addition, the energy spectra of charged secondaries are needed to study the radiation signal exiting the patient accounting for tissue inhomogeneities, location of the tumour, treatment plan parameters and performance of monitoring detector.
 
The experimental setup used for measurements and data selection of the analysis are described in Sections~\ref{sec:expSetup} and~\ref{sec:dataSelectionAndPid}. Three crucial properties of the charged secondaries production were investigated: Section~\ref{sec:flux} focuses on the yield of the secondary protons, Section~\ref{sec:eneSpectra} on proton energy spectra, and Section~\ref{sec:profiles} on proton emission profiles. For each primary ion beam the measured charged secondary emission profile was related to the expected dose deposition profile in order to investigate the feasibility of a dose monitoring technique in PT. The angular dependence of the charged secondary radiation emission was studied at 60\degree ~and 90\degree ~with respect to the primary beam. The experiment was performed in 2014 at Heidelberg Ion Beam Therapy (HIT) centre, Germany, a hospital based facility using proton and carbon beams for patient treatment since 2009. 

\section{Experimental setup}
\label{sec:expSetup}
\begin{figure}
\centering{} 
\includegraphics[width=0.51\textwidth]{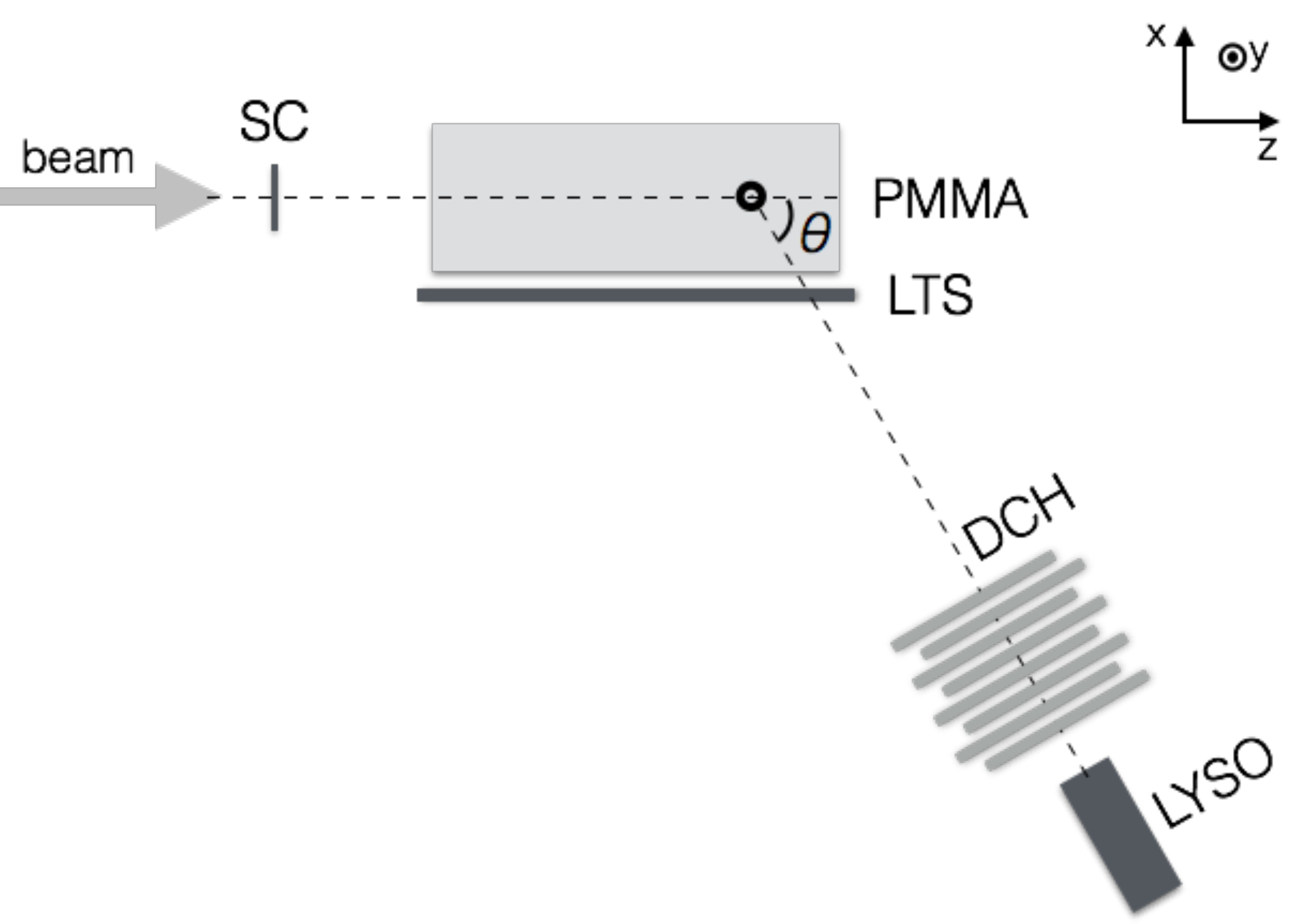} 
\caption{Experimental setup used for the measurement of charged secondary products generated by primary beam impinging the PMMA target (not to scale).
DCH and LYSO detectors were mounted on a movable arm situated for different measurements at $\theta$=60\degree or $\theta$=90\degree with respect to the primary beam direction. The origin of the reference frame is marked by the black spot inside the PMMA target,  $\sim$1\,cm before the distal edge of PMMA box.}
\label{fig:expSetup}
\end{figure}

The measurements were performed in the HIT experimental cave. The secondary radiation was detected at 60\degree~and 90\degree~with respect to the primary \he~ ion beam impinging on the PMMA target, and at 90\degree ~with respect to \car~ ion beam (see Fig.~\ref{fig:expSetup},~Tab.~\ref{tab:beams}). A constant PMMA length along the beam (d$\rm _{PMMA}$) was used for \car~ion runs, whereas for \he~ ion runs this length was adjusted according to the primary beam energy. 
The reference frame is depicted in the Fig.~\ref{fig:expSetup}: beam direction is referred as z, whereas x and y define the transverse plane with respect to the beam. The complete geometry of the experimental setup was implemented in the FLUKA \cite{Ferrari2005,Boehlen2014} Monte Carlo (MC) code to simulate and study detector acceptance, efficiency and particle identification.
 
The PMMA target ($\rm 5 \times 5$\,cm$^{2}$ face orthogonal to the beam line, density $\rm 1.19\,g\cdot cm^{-3}$, ionization potential 74\,eV) was positioned at the beam isocenter $\sim$1 m away from the beam nozzle and with its longer side $\rm d_{PMMA}$ along the beam line (Fig.~\ref{fig:expSetup}). 
The charged secondaries produced in the PMMA had to travel on average 2.5\,cm of material to exit the target, in the 90\degree ~setup configuration. A pencil beam with Gaussian spot size was used, with Full Width at Half Maximum (FWHM) ranging from 4.7 to 9.3 mm depending on the beam and its energy (see Tab.~\ref{tab:beams}).  For each \he ~primary beam energy, the PMMA target length $\rm d_{PMMA}$ along the beam was selected to keep the position of the BP inside the PMMA, before its exit face, as indicated in Fig.~\ref{fig:expSetup}. The \car~ion beam at 220\,MeV/u stopped close to the end of 10\,cm-long PMMA target, whereas less energetic \car ~ion beams were stopped earlier in the PMMA target, on the line between PMMA entrance face and the origin of the reference frame. Tab.~\ref{tab:beams} lists the primary beam energy, range in PMMA (computed by FLUKA MC simulations) and transverse size (FWHM) as well as the  d$\rm _{PMMA}$ PMMA length used in the experiment.

The number of primary ions (primary beam rate) impinging on the PMMA target was measured using a 0.2\,cm-thick plastic scintillator (Start Counter - SC; Fig.\,\ref{fig:expSetup}) positioned upstream at 37\,cm from the PMMA target and read out by two opposite photomultiplier tubes (PMT; Hamamatsu H10580). The angular distribution of the secondary particles produced in the target were studied at 90\degree~and 60\degree~with respect to the primary beam. For this purpose three isocentrically positioned detectors were mounted on a movable arm: 0.1\,cm-thick plastic scintillator (LTS), 21\,cm-long drift chamber (DCH) and a matrix of four cerium doped lutetium-yttrium oxyorthosilicate crystals (LYSO), 1.5\,x\,1.5\,x\,12\,cm{$^3$} each. The position of LTS, DCH and LYSO detector front faces, with respect to PMMA central axis, were 8.0\,cm, 50.5\,cm (90\degree) and 55.0\,cm (60\degree), 73.5\,cm (90\degree) and 78.0\,cm (60\degree) respectively. The scintillation light of LYSO crystals was detected with a Photomultiplier Tube (EMI 9814B PMT). The response of LYSO crystals was evaluated with the HIT accelerator proton beam. The crystal matrix was centered in front of the beam nozzle, parallel to the beam, exposing the four crystals to the same average proton yield. The LYSO matrix was irradiated with proton beams of seven energies in 50--200\,MeV range. Four LYSO crystals showed a different light yield response that was taken into account in the particle identification.

The production point of the charged secondaries was obtained by three dimensional reconstruction of the particle track with a drift chamber (DCH)~\cite{AbouHaidar2012}, consisting of six alternated horizontal (x-z plane) and six vertical (y-z plane) wire layers. The DCH was operated applying high voltage of 1.8 kV to the sense wires and flushing the active volume with an Ar/CO$_2$ (80/20) gas mixture, as described in~\cite{Piersanti2014}. The output signals were discriminated applying a 30\,mV threshold. In this configuration the single cell spatial resolution is 200\,$\mu$m and the single cell efficiency is $\simeq$96\% (Abou-Haidar et al 2012). The readout and performances of the DCH as well as the tracking algorithm and DCH calibration procedure can be found elsewhere (Agodi et al 2012b).

The triggering logic implemented for the selection of charged secondaries required the SC and LYSO signals coincidence within 80 ns time window. The front-end electronics, used to acquire time and charge information from all above described detectors, was read out by a VME system interfaced with a PC Data AcQuisition (DAQ) server, as it was described in details elsewhere~\cite{Piersanti2014}. At the highest delivered beam rate of $\sim$3\,MHz, the trigger rate was in 0.3\,--\,6\,kHz range.

\begin{table}[ht!]
\centering
\caption{Beam and setup properties used in the measurements; B$_{\rm FWHM}$ - spot size, beam Range in PMMA, $\rm d_{PMMA}$ - PMMA length along the beam, $\theta$ - detector position with respect to the primary beam direction.}
\label{tab:beams}
{
\renewcommand\arraystretch{1.1}
\centering
\begin{tabular}{c c c c c c}
\textbf{Ion} & \textbf{Energy} & \textbf{B$_{\rm \textbf{FWHM}}$ } & \textbf{Range} & \textbf{d$_{\rm \textbf{PMMA}}$}
& $\bm{ \theta}$ \\
&  (MeV/u) & (mm) & (cm) & (cm) &  \\ \hline  
\multirow{4}{*}{$^{12}{\rm C}$} & $120$ & 7.9 & 2.9 & \multirow{4}{*}{10.0} & \multirow{4}{*}{$90\degree$}\\ 
 & $160$ & 6.2 & 4.8 & \\
& $180$ & 5.5 & 6.0 & \\
& $220$ & 4.7 & 8.3 & \\ \hline
\multirow{3}{*}{$^{4}{\rm He}$} & 102 & 9.3 & 6.7 & 7.7 & $60\degree$ \\ \cline{6-6}
 & $125$ & 7.8 & $9.7$ & $10.0$ & \multirow{2}*{$90\degree$\ -\ $60\degree$} \\ 
& $145$ & 6.9  & 12.5 & 12.7 & \\ \hline
\end{tabular}
}
\end{table}

\section{Data selection and particle identification}
\label{sec:dataSelectionAndPid}

\begin{figure}
\centering{}
\subfigure[\car, 90\degree ~configuration] {\includegraphics[trim=0.cm 0.cm 0.cm 0.cm, clip,  width=0.32\textwidth]{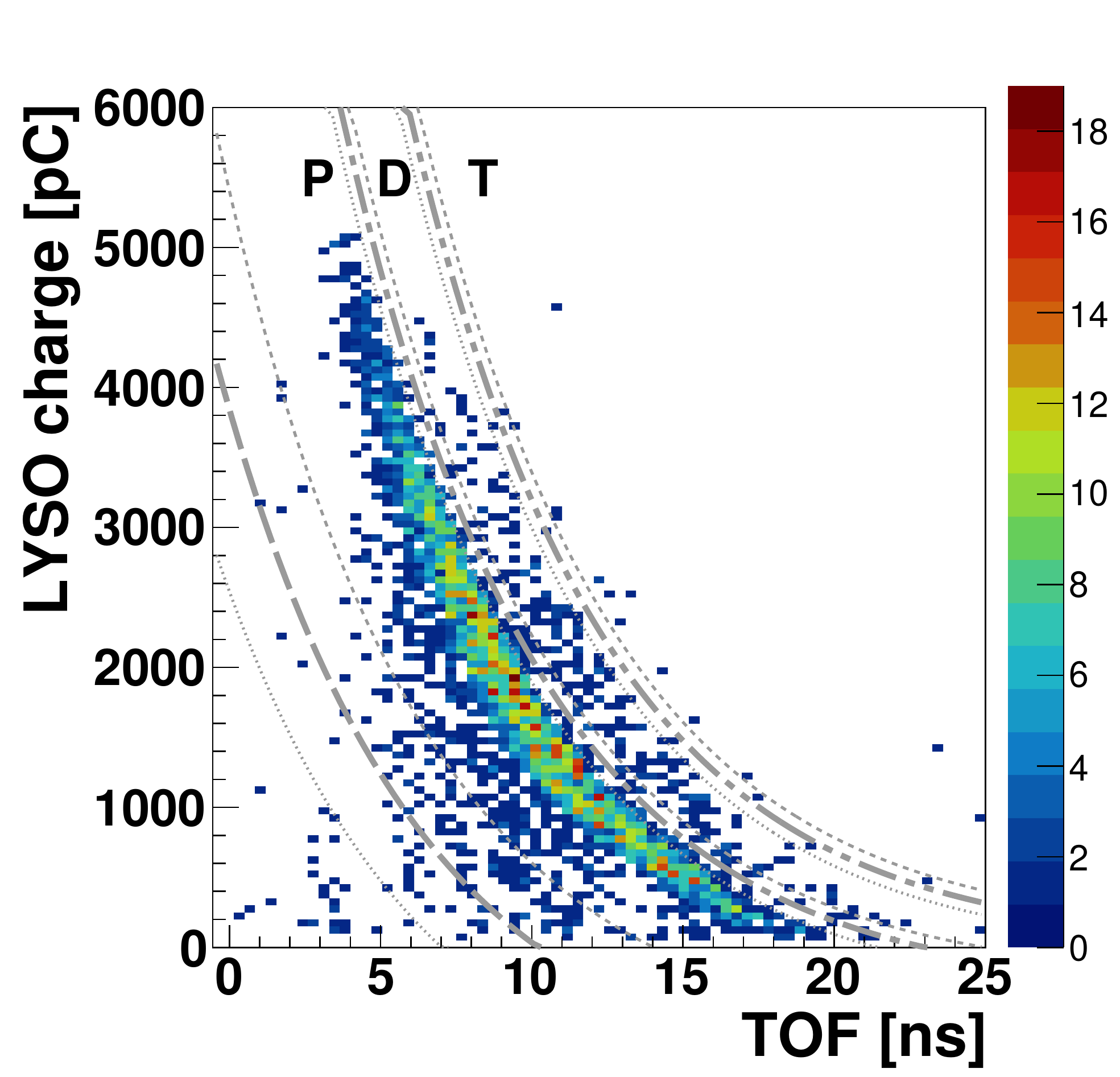}}
\subfigure[\he, 90\degree ~configuration] {\includegraphics[trim=0.cm 0.cm 0.cm 0.cm, clip, width=0.32\textwidth]{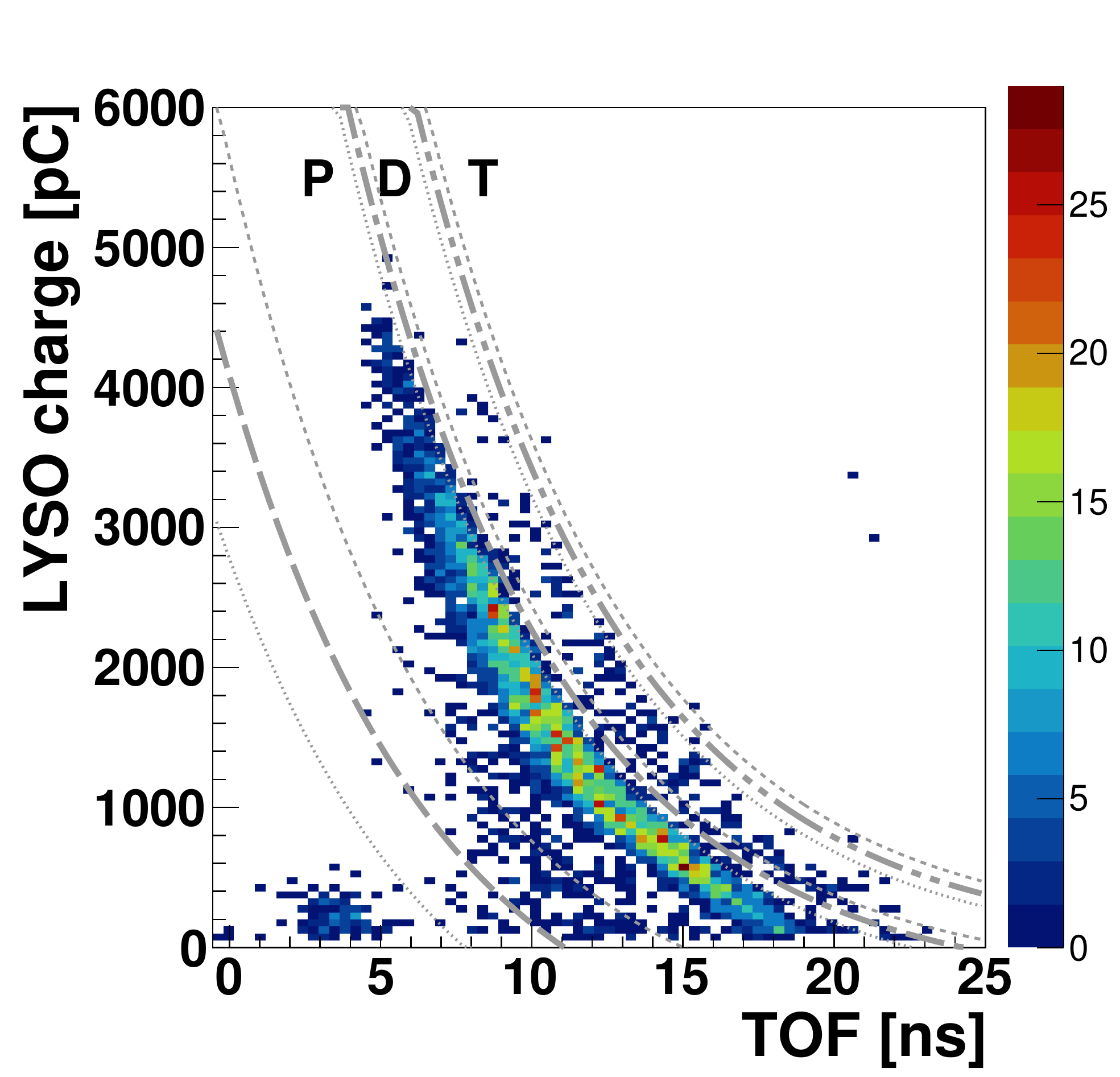}}
    \subfigure[\he, 60\degree ~configuration] {\includegraphics[trim=0.cm 0.cm 0.cm 0.cm, clip, width=0.32\textwidth]{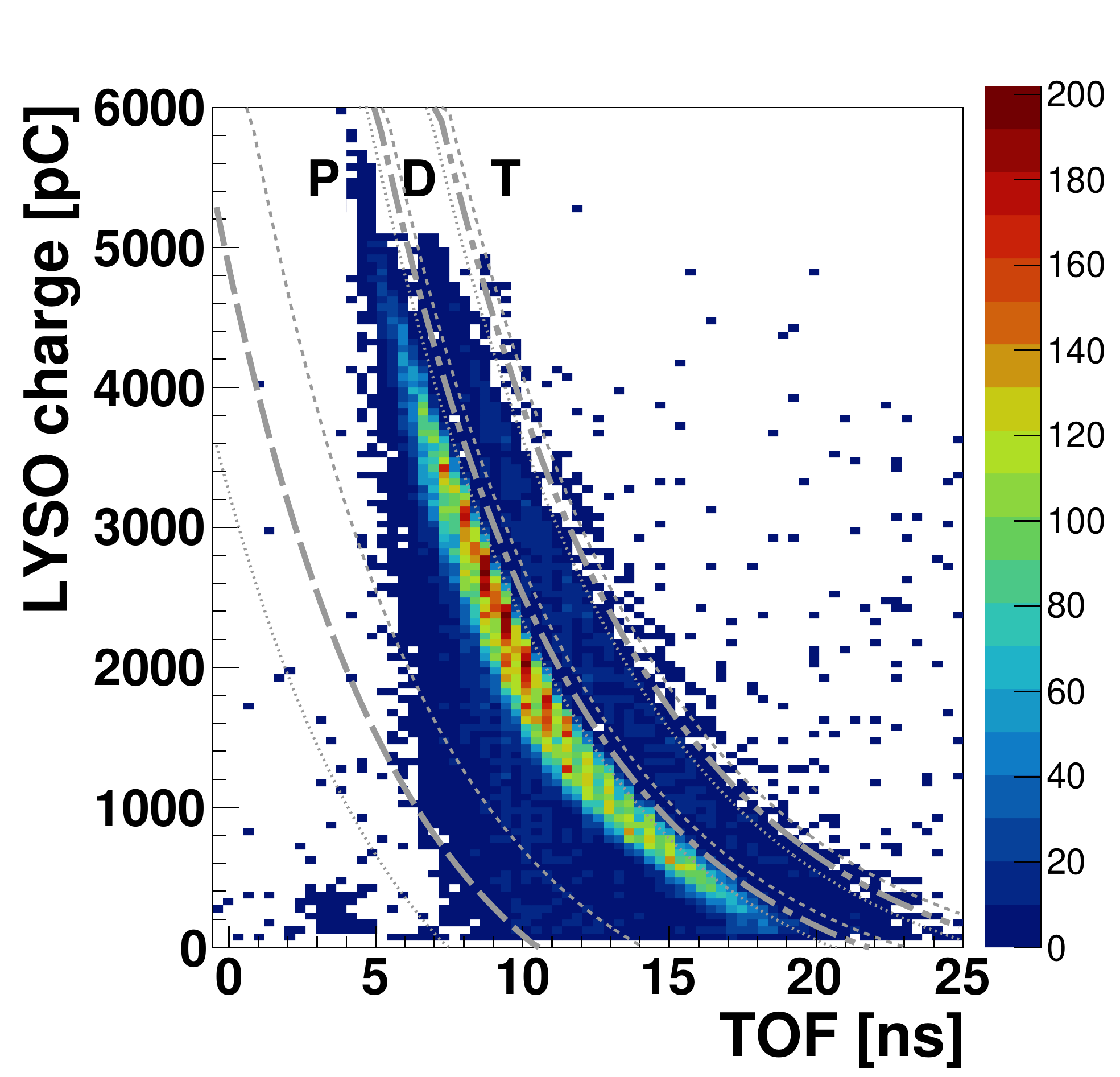}}
    \caption{The number of secondary protons (raw data) produced in the PMMA as a function of TOF and charge produced by LYSO detector. Protons (P), Deuterons (D) and Tritons (T) can be distinguished. The population of electrons with TOF~$\sim$3.5\,ns and energies up to few MeV/u was excluded by PID. Dashed, dashed-dotted, dashed-dotted-dotted lines show how the P, D, and T were identified, respectively. PID was used to calculate P, D, and T yields and energy spectra. Shifting the PID lines the systematic uncertainty on the yield was estimated. }
\label{fig:qLyso_vs_TOF}
\end{figure}

The selection of charged secondaries was performed by exploiting the DCH information together with the energy released in the LYSO detector and the Time of Flight (TOF) defined as the time difference between LTS and LYSO detector signals. Most of the events with charged particles in the final state fire N$_{\rm{DCH}}$=12 DCH cells, one cell in each DCH plane. In order to classify an event as given by charged secondaries, N$_{\rm DCH}\geq$\,8 was required. Fig.~\ref{fig:qLyso_vs_TOF} illustrates the number of events as a function of charge produced by LYSO detector and TOF. All the events collected with the Carbon ion beam and the 90\degree\ configuration (a) as well as Helium ion beam with 90\degree\ (b) and 60\degree\ (c) setup configuration (N$_{\rm DCH}\geq$\,8; all the investigated energies) are shown. Three bands characteristic of Proton (P), Deuteron (D) and Triton (T) events are visible. The population of events with TOF~$\sim$3.5\,ns and energies up to few MeV/u was identified as electrons. Such a result is confirmed by the data/MC comparison.

Particle identification (PID) was performed using the selection bands for proton, deuteron and triton populations as indicated in Fig.~\ref{fig:qLyso_vs_TOF} (bold lines). The deuteron contribution is 5\% and 10\% of all events (P+D+T) detected at 90\degree ~and 60\degree ~respectively, whereas the triton contribution is at the level of 1-2\% in all cases.
In order to account for the underlying background contribution from neighboring populations (e.g. deuteron background in proton distribution), the P-D and D-T separation lines were moved and the PID systematic uncertainty on the yields were estimated. Distributions shown in Fig.~\ref{fig:qLyso_vs_TOF} were obtained with number of primary ions $\rm N_{ion}=3.5\times 10^{9}$, $\rm N_{ion}=7.2\times 10^{9}$ and $\rm N_{ion}=6.7\times 10^{9}$ for (a), (b) and (c) configurations respectively. For the three different configurations a total number of secondary particles (P+D+T) equal to 3753, 4676 and 51711 was measured. 

\section{Yield and efficiency evaluation}
\label{sec:flux}

\begin{figure}[t]
 \centering{}
\subfigure[$\rm N_{p}(E_{kin}^{Det},z)$ at 90\degree] 
{\includegraphics[width=0.31\textwidth]{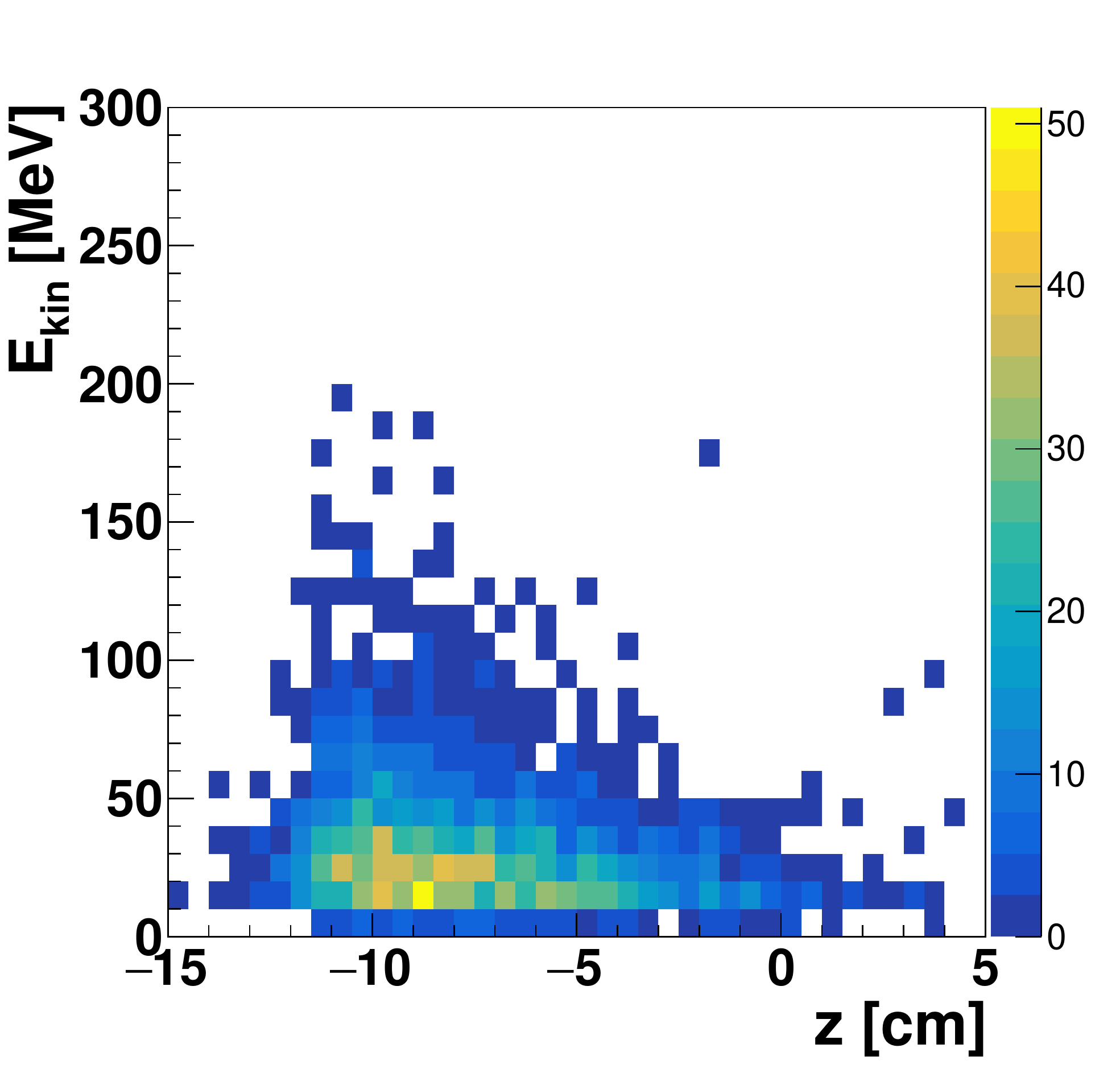}}
\subfigure[$\rm \epsilon_{p} (E_{kin}^{Det},z)$ at 90\degree] {\includegraphics[width=0.31\textwidth]{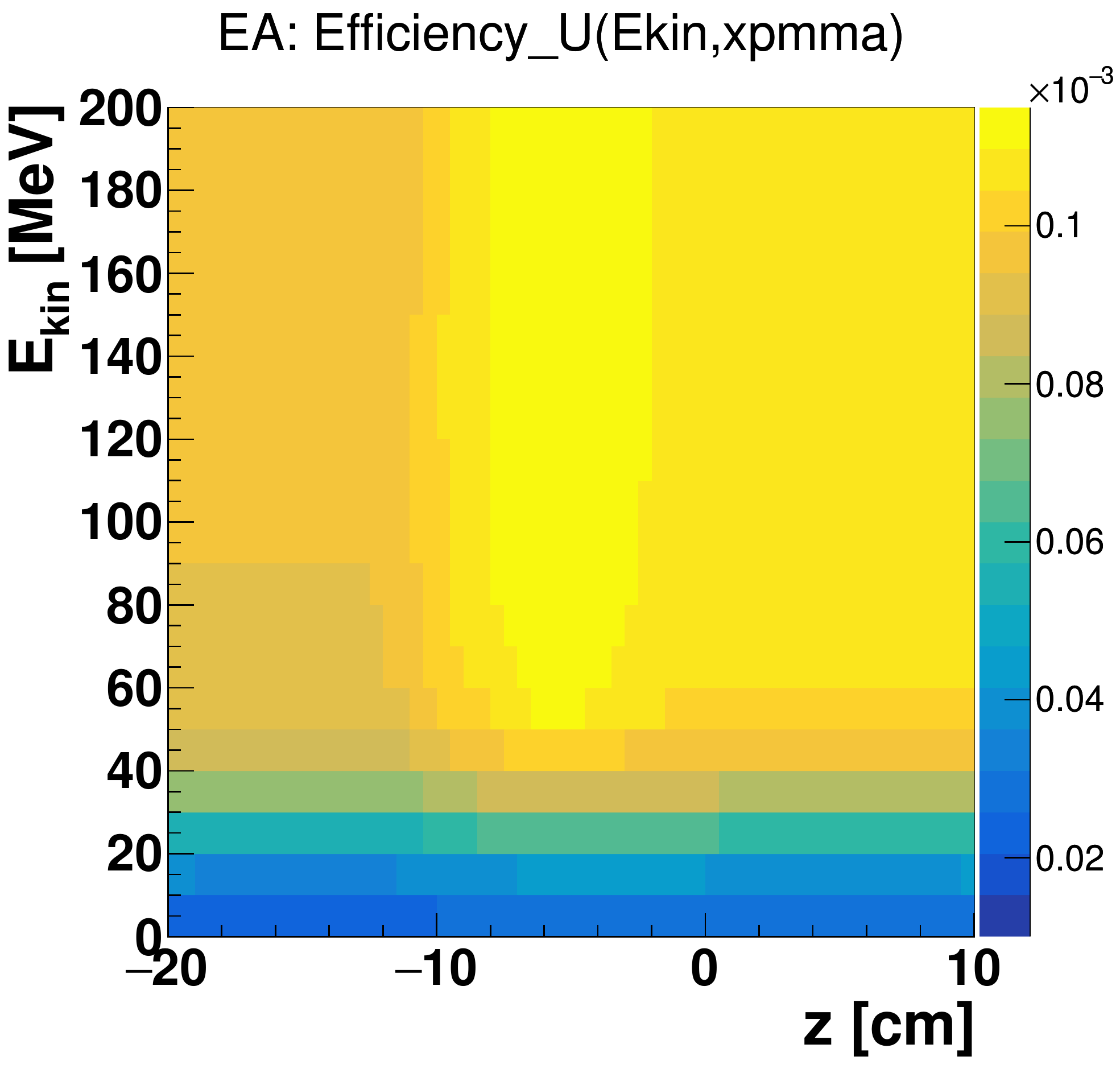}}
\subfigure[$\rm \Phi_{p}(E_{kin}^{Det},z)$ at 90\degree]
{\includegraphics[width=0.31\textwidth]{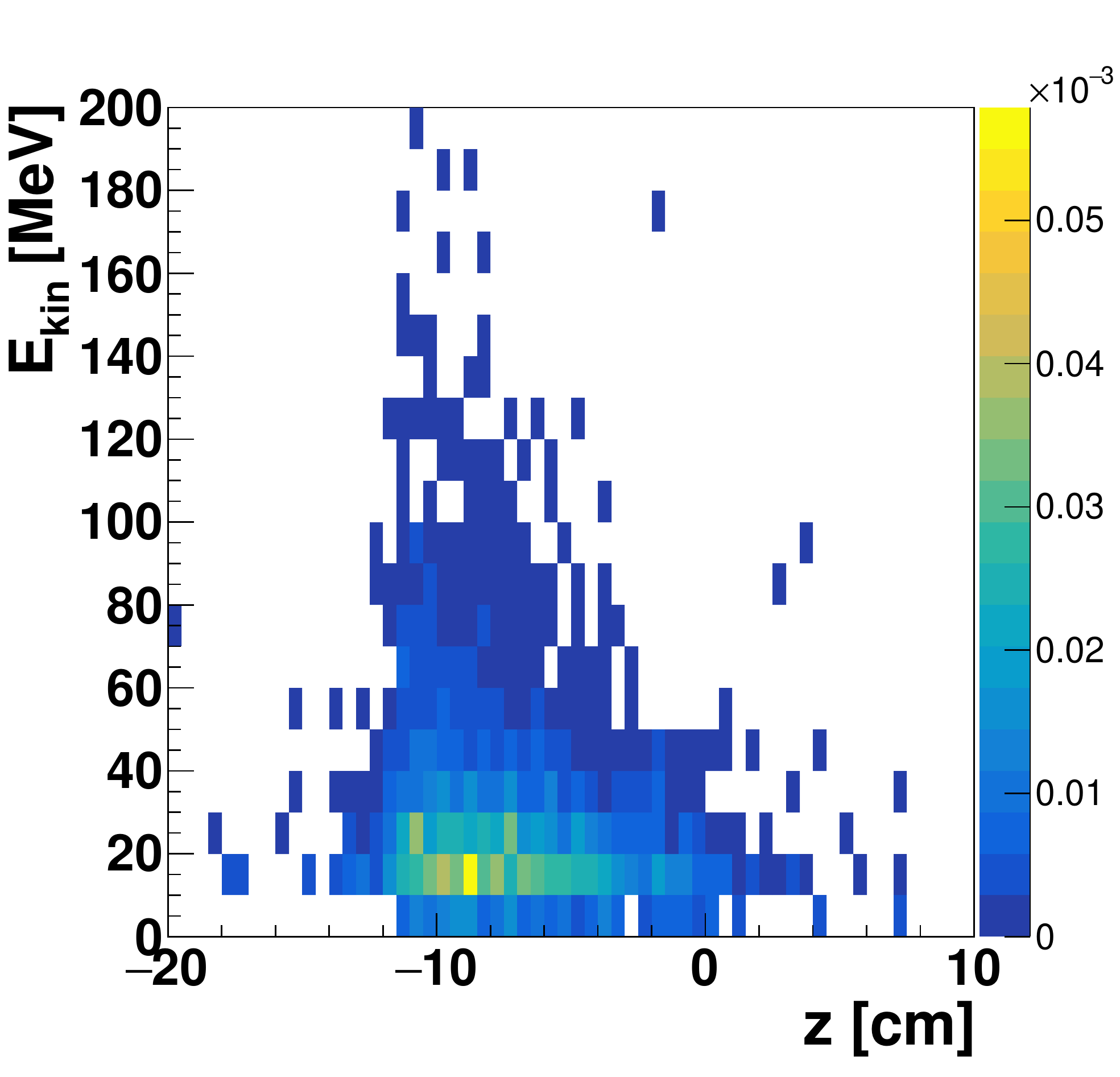}}
\subfigure[$\rm N_{p}(E_{kin}^{Det},z)$ at 60\degree] {\includegraphics[width=0.31\textwidth]{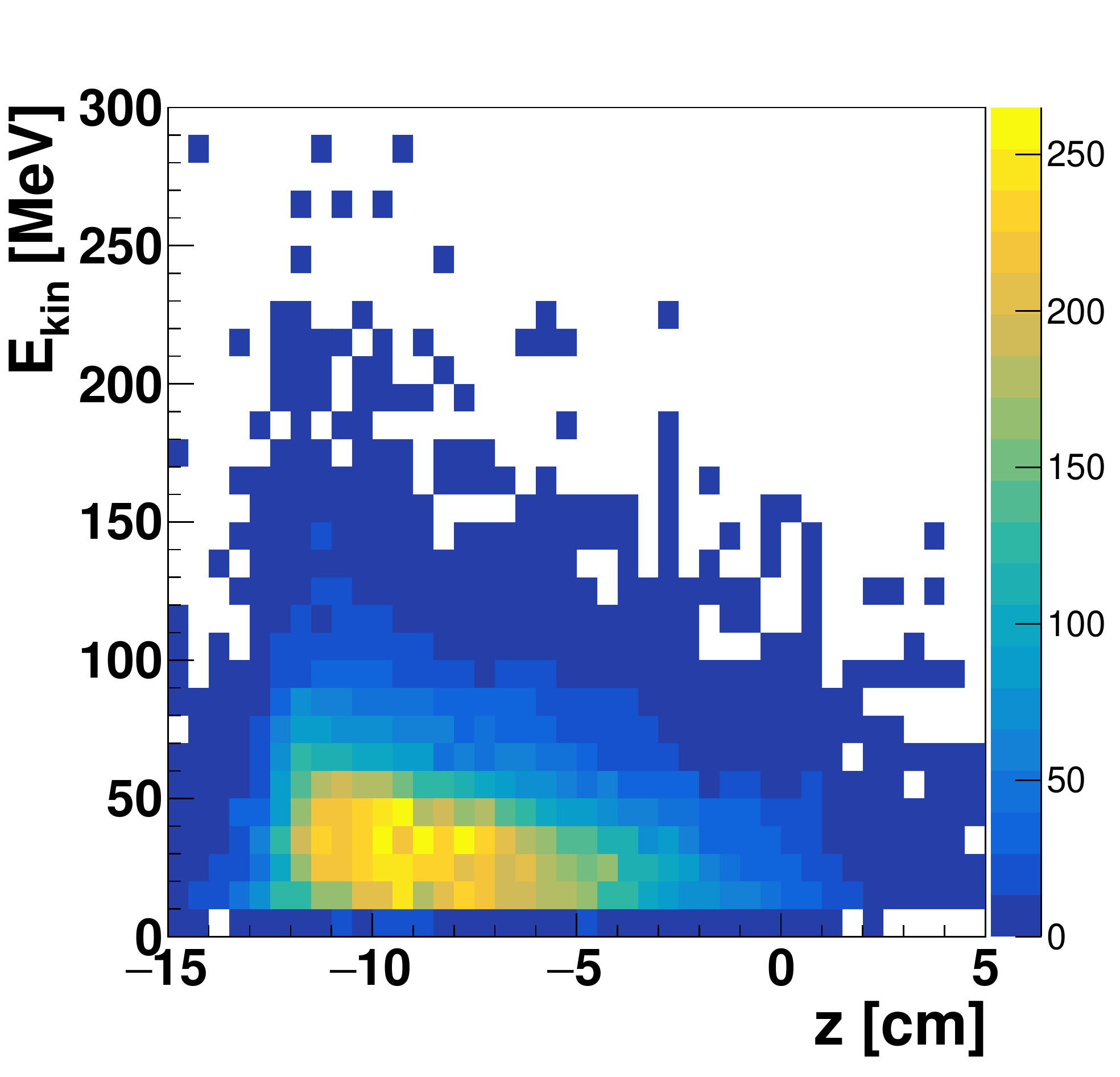}} 
\subfigure[$\rm \epsilon_{p} (E_{kin}^{Det},z)$ at 60\degree] {\includegraphics[width=0.31\textwidth]{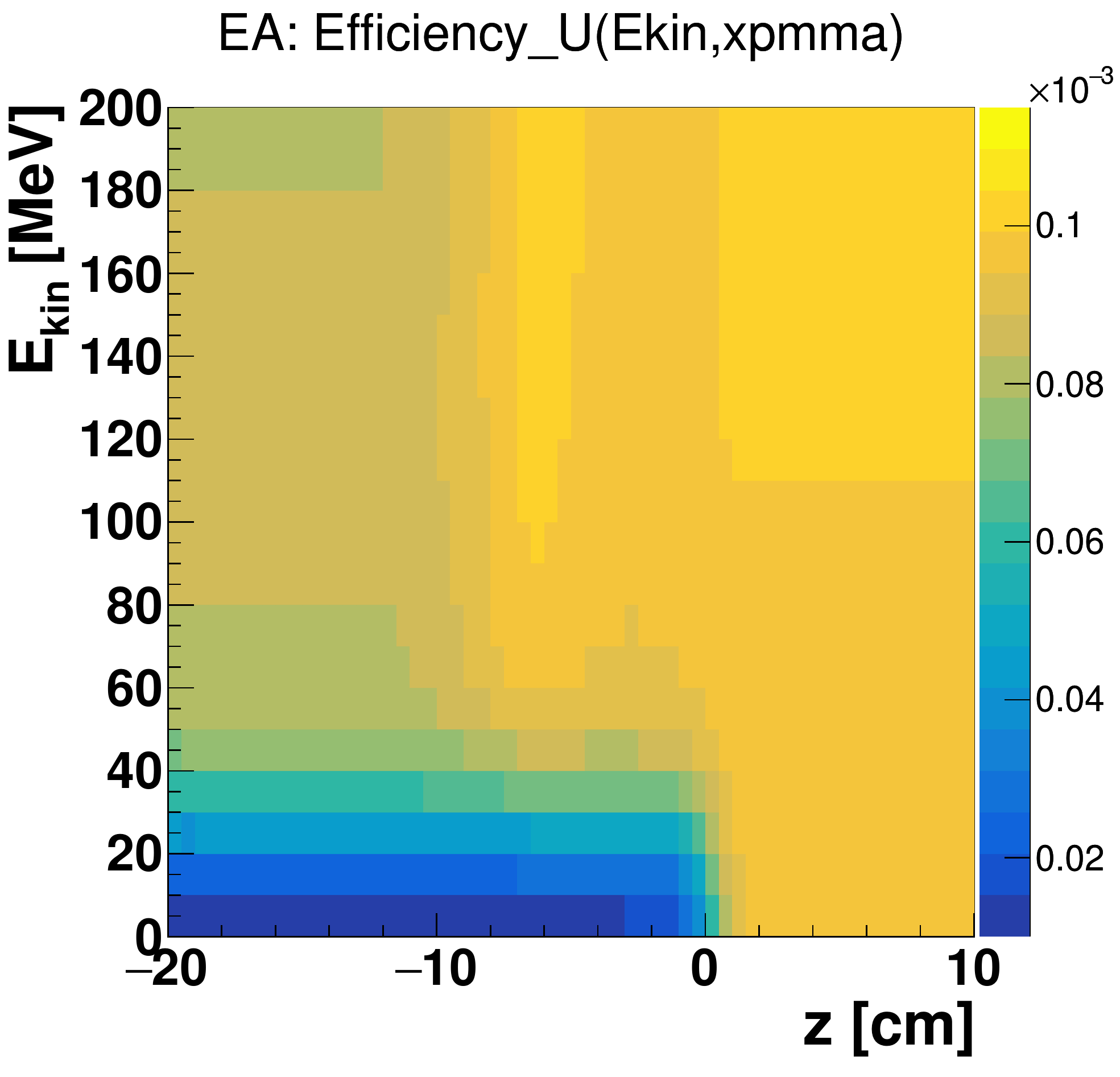}}
\subfigure[$\rm \Phi_{p}(E_{kin}^{Det},z)$ at 60\degree] {\includegraphics[width=0.31\textwidth]{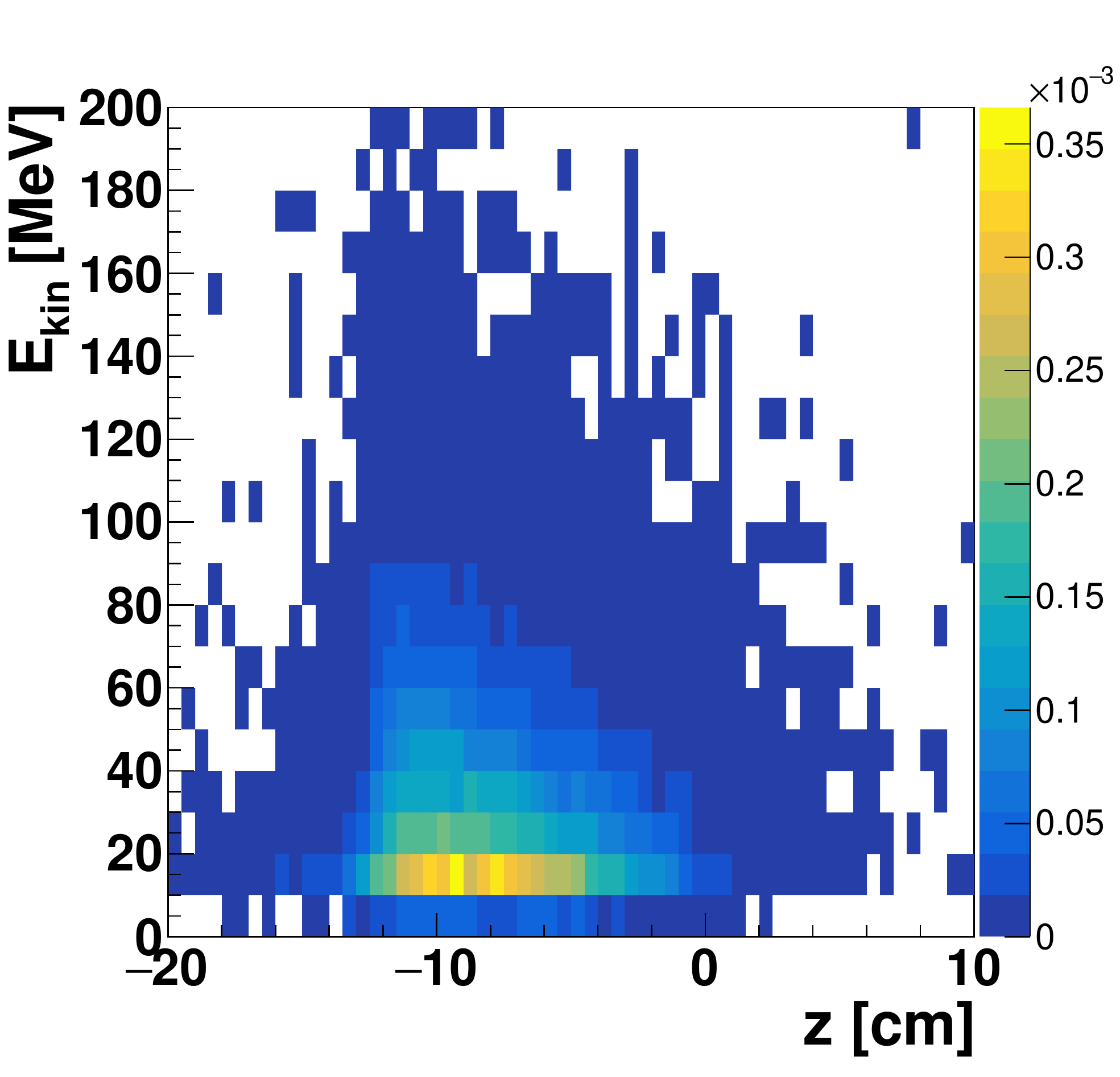}}
\caption{Data (a,d): the number of secondary protons (raw data) produced by \he~beam at 145~MeV/u. MC (b,e): the efficiency maps obtained from MC simulations for secondary protons. Data corrected (c,f): $\rm \Phi_{p}(E_{kin}^{Det},z)$ - the secondary proton yield for \he~beam at 145~MeV/u. The data are plotted as a function of the detected kinetic energy ($\rm E_{kin}^{Det}$) and reconstructed production point (z). Top (a,b,c) and bottom (d,e,f) plots illustrate the results obtained with detector positioned at 90\degree ~and 60\degree ~with respect to the primary beam, respectively.
The energy spectra reported in section~\ref{sec:eneSpectra} were built from a profile of the right plot (c,f) on y-axes. The emission shapes reported in section~\ref{sec:profiles} were built from a profile of the left plot (a,d) on x-axes. }
    \label{fig:dataAndeffMap}
  \end{figure}

The differential production rate of charged secondary particles, normalized to the number of primary ions, averaged on the total solid angle and integrated over the full target length (i.e. yield) was estimated, for Helium and Carbon ion beams, as:
\begin{equation}\rm \Phi_{p}=\frac{dN_{p}}{N_{ion}d\Omega}=\frac{1}{4\pi}\frac{1}{N_{ion}\epsilon_{DT}}\sum_{E_{kin}^{Det}}\sum_{z}\frac{N_{p}(E_{kin}^{Det},z)}{\epsilon_{p}(E_{kin}^{Det},z)}   , 
\label{eq:eff_def}
\end{equation}
where $\rm N_{p}(E_{kin}^{Det},z)$ is the number of detected protons, $\rm N_{ion}$ is the number of primary ions impinging on the PMMA target, $\rm \epsilon_{\rm DT}$ is the dead time (DT) efficiency, and $\rm \epsilon_{p}(E_{kin}^{Det},z)$ is the total detection efficiency computed as a function of the production point (z) and of the kinetic energy ($\rm E_{kin}^{Det}$) of the secondary particles. TOF between LTS and LYSO detectors was used to measure the $\rm E_{kin}^{Det}$ of detected secondary particles whereas their production point in PMMA was reconstructed using the DCH information.

The total number of primary ions impinging on the PMMA target $\rm N_{ion}$ is computed counting the number of SC signals and correcting it for the dead time efficiency introduced by the discrimination time of the trigger signals. The correction factor and its systematic uncertainty was obtained specifically for each run from the dedicated MC simulations and ranges from 1.03 to 1.54, as described in details in \cite{Mattei2016}.

DT efficiency was evaluated using the VME system, counting all the generated trigger signals ($\rm N_{TrTot}$) and the triggers signals acquired by the DAQ system ($\rm N_{TrAcq}$). This DT efficiency, defined as $\rm \epsilon_{\rm DT} = N_{TrAcq}/N_{TrTot}$, varied from 60\% to 90\%, depending on the beam rate. Run specific values of $\rm \epsilon_{\rm DT}$ were used to compute the yield using Eq.\,1 for the different data acquisition conditions (primary ion beam, beam energy and angular configuration).

The total proton detection efficiency $\rm \epsilon_{p}(E_{kin}^{Det},z)$ including detector efficiencies (LTS, DCH, LYSO) was computed from a dedicated MC simulation accounting for the complete setup geometry. $\rm \epsilon_{p}(E_{kin}^{Det},z)$ varies as a function of the production point (z) of secondary particles, due to the geometrical acceptance of the DCH-LYSO system and as a function of the kinetic energy ($\rm E_{kin}^{Det}$) of secondary particles. This last dependency is primarily due to the energy lost to escape the PMMA and to the primary beam spot size in the transverse plane.

The total detection efficiency map $\rm \epsilon_{p}(E_{kin}^{Det},z)$ obtained from MC simulation is shown on Fig.\,\ref{fig:dataAndeffMap}~(b,e) for 90\degree ~and 60\degree ~setup configuration, respectively. Due to the beam spot size of FWHM up to $\sim$1\,cm, the secondary particles have to travel 2-3 \,cm within the PMMA before exiting the target. Therefore the detection efficiency of the particles with the lower $\rm E_{kin}^{Det}$ values (0-40\,MeV; blue area in Fig.\,\ref{fig:dataAndeffMap}:\,b,\,e) is smaller than the detection efficiency of the more energetic particles (the minimal $\rm E_{kin}^{prod}$ needed to exit the PMMA depends mainly on particle's production point in the x direction). The minimal production energy of protons needed to exit PMMA was estimated from MC simulation to be about $\rm E_{kin}^{prod}=50$\,MeV (cf. Sec.\,\ref{sec:eneSpectra} and Fig.\,\ref{fig:EkinProd_vs_EkinDet}). 
The efficiency map $\rm \epsilon_{p}(E_{kin}^{Det},z)$ was built based on the efficiency calculation performed using 10 energies in the range $\rm E_{kin}^{prod}$=50-250\,MeV and using production points uniformly distributed along z. The obtained distribution was then smoothed to provide the efficiency values for secondaries at 10\,MeV energy steps and accounting for their production position in PMMA along the $\rm z$ axis in 5~mm steps (coordinate system introduced in Fig.~\ref{fig:expSetup}). 

The proton yield ($\rm \Phi_{p}$) over the detection threshold ($\rm E_{kin}^{prod}>50$\,MeV) was obtained from the number of detected protons ($\rm N_{p}$) as a function of the production point (z) and the detected kinetic energy ($\rm E_{kin}^{Det}$), as illustrated in Fig.~\ref{fig:dataAndeffMap}. The number of measured events in each bin ($\rm N_{p}(E_{kin}^{Det},z)$, Fig.~\ref{fig:dataAndeffMap}\,a,d) was corrected for the efficiency ($\rm \epsilon_{p}(E_{kin}^{Det},z)$; Fig.~\ref{fig:dataAndeffMap}\,b,e) providing the yield for each bin of kinetic energy ($\rm E_{kin}^{Det}$) and production point (z) ($\rm \Phi_{p}(E_{kin}^{Det},z)$, Fig.~\ref{fig:dataAndeffMap}\,c,f). The integrated number of events corrected for total detection efficiency, dead time efficiency and normalized to the number of primaries shown in Fig.~\ref{fig:dataAndeffMap}\,c,f corresponds to the proton yield ($\rm \Phi_{p}$) given in Tab.~\ref{tab:fluxesP}.

\begin{figure}[t]
\centering\includegraphics[width=.4\linewidth, trim={.3cm .cm 1.cm 1.4cm},clip]{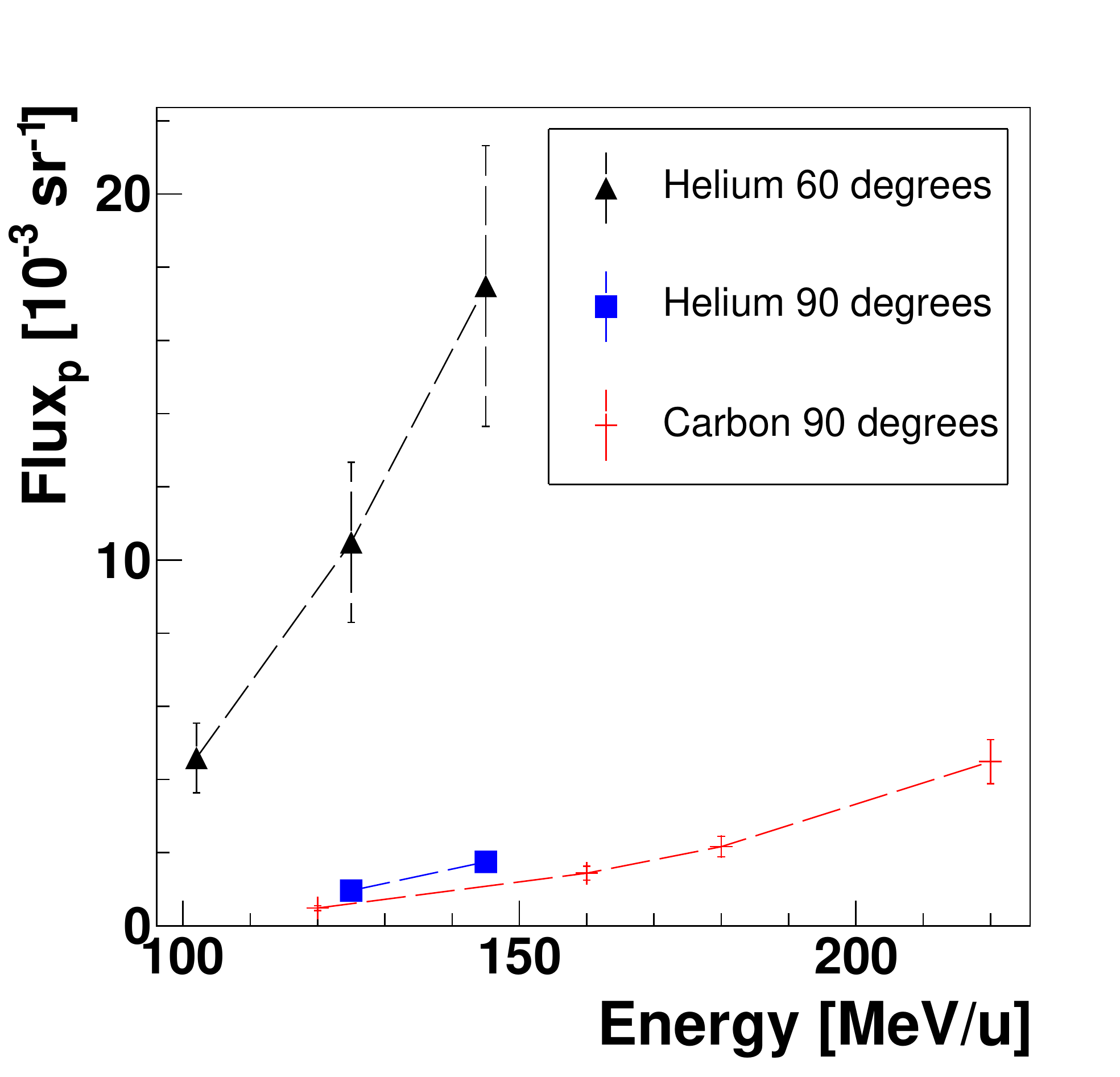}\hfill
\caption{ Secondary proton yields ($\rm \Phi_{p}$) obtained with \car~ and \he ~beams using 90\degree ~and 60\degree ~setup configuration plotted as a function of kinetic energy per nucleon of primary beam (Energy). Error bars correspond to the statistical and systematic uncertainties summed in quadrature.}
\par\vspace{0pt}
\centering%
\label{fig:fluxes}
\end{figure}

Fig.~\ref{fig:fluxes} shows $\rm \Phi_{p}$ as a function of \car ~and \he ~ion beam energy for the detector positioned at 90\degree ~and 60\degree ~with respect to the primary beam direction. The measured yields are reported with both statistical and systematic uncertainties in Tab.~\ref{tab:fluxesP}. The number of secondary particles produced in the target increases with the energy of the primary beam, i.e. with its range. Comparing ion beams having a similar range, the yield produced by the \car ~ion beam at 220\,MeV/u is higher than the one produced by \he ~beam at 125\,MeV/u, as the secondary particles are produced essentially in projectile fragmentation. The secondary particle yield induced by \he ~ion beam of a similar range and detected at 60\degree ~with respect to the primary beam direction is one order of magnitude higher than secondary particle yield detected at 90\degree. 

\begin{table}
\centering
\caption{Yields of secondary protons ($\rm \Phi_{p}$) for Carbon and Helium primary beams. The table includes information about the primary ion beam (Beam), its kinetic energy per nucleon (Energy) and setup configuration ($\rm \theta$) used for the measurement.}
\label{tab:fluxesP}
\vspace{0.2cm}
 \begin{tabular}{cccc}
$\rm \mathbf{\theta}$ & \textbf{Beam} & \textbf{Energy} &  $\rm \mathbf{ \Phi_{p} \pm \sigma _{(stat)} \pm \sigma _{(sys)}}$\\
  &  & [MeV/u] & [$10^{-3} sr^{-1}$]\\ \hline
\multirow{6}{*}{90\degree} & \multirow{4}{*}{\car} & 120 & 0.5 $ \pm 0.0 \pm 0.1 $ \\
 & & 160 & $1.4 \pm 0.1 \pm 0.2 $ \\
 & & 180 & $2.2 \pm 0.1 \pm 0.3 $ \\
 & & 220 & $4.5 \pm 0.1 \pm 0.6 $ \\ \cline{2-4}
& \multirow{2}{*}{\he} & 125 & $1.0 \pm 0.0 \pm 0.1 $ \\
& & 145 & $1.7 \pm 0.0 \pm 0.2 $ \\ \hline
\multirow{3}{*}{60\degree} & \multirow{3}{*}{\he} & 102 & $4.6 \pm 0.1 \pm 1.0 $ \\
& & 125 & $10.5 \pm 0.1 \pm 2.2 $ \\
& & 145 & $17.5 \pm 0.1 \pm 3.8 $ \\ \hline
\end{tabular} 
\end{table}

\vspace{0.3cm}

The total uncertainty on the yield consists of both statistical and systematic contributions. Fractional statistical uncertainty ranges from 1\% to 7\% depending on the primary ion beam, its energy and setup configuration (90\degree~or~60\degree) and is mainly due to the statistical uncertainty on the number of detected charged secondary particles. The fractional statistical uncertainty contribution from both the detection efficiency (due to the MC sample statistics) and the number of primary ions is at few per mil level. 

The fractional systematic contribution generally dominates the yield uncertainty and ranges from 12\% to 22\%. The main fractional uncertainty contribution comes from the efficiency map estimation method and ranges from 8\% to 19\%. Systematic uncertainty related to the correction to the total number of primary ions \cite{Mattei2016}, estimated from a dedicated MC simulation is a function of the primary ion beam rate and it is within the 2\%-7\% range. The systematic uncertainty related to the calculation of the raw number of the primary ions has been assessed with an independent ions counting method using the external PET detectors available during the data taking as described in details elsewhere~\cite{Mattei2016} and ranges from 4\% to 6\%. The systematic uncertainty for PID (Fig.\,\ref{fig:qLyso_vs_TOF}) ranges from 3\% to 6\%. The systematic uncertainty related to the less and more rigorous DCH selection criteria (N$_{\rm{DCH}}\geq$7 or N$_{\rm{DCH}}\geq$9) is negligible. The contribution from the dead time correction is also negligible.

The yield obtained with primary Carbon ion beam at 220\,MeV/u and applying the same selection criteria as in~\cite{Piersanti2014} is $(2.8 \pm 0.1_{(stat)} \pm 0.2_{(sys)}) \times 10^{-3} sr^{-1}$. This result is in agreement within uncertainties with the total yield $(2.7 \pm 0.0_{(stat)} \pm 0.1_{(sys)}) \times 10^{-3} sr^{-1}$ obtained by ~\cite{Piersanti2014} . 
\section{Energy spectra}
\label{sec:eneSpectra}

\begin{figure}[t]
\centering\includegraphics[width=.4\linewidth, trim={.3cm .cm 0.cm 0.cm},clip]{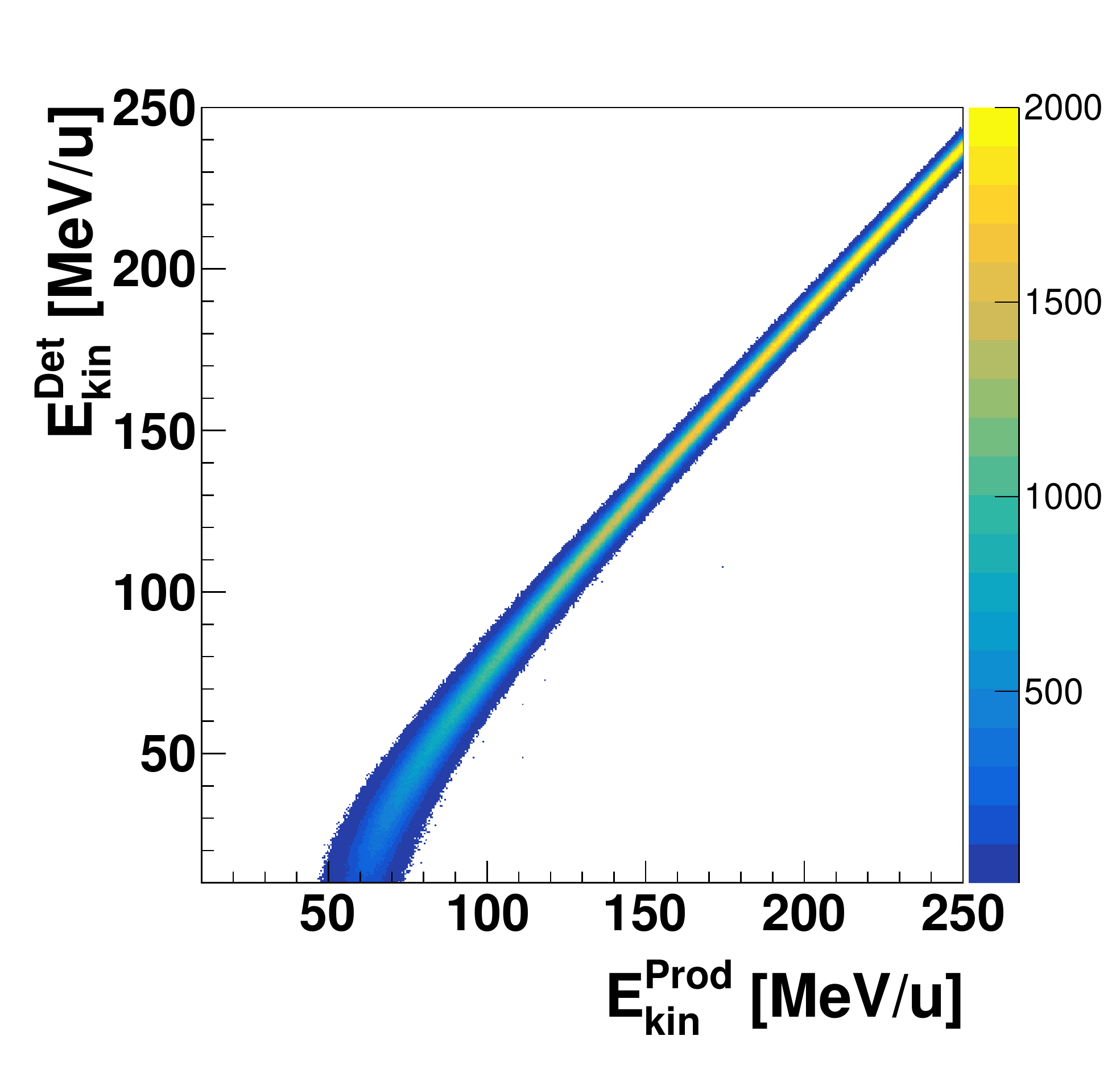}\hfill
\caption{The relation of kinetic energy of secondary protons detected by LYSO ($\rm E_{kin}^{Det}$) plotted versus kinetic energy of secondary protons at production point inside PMMA ($\rm E_{kin}^{Prod}$) obtained from high statistics MC simulations. Color scale corresponds to the number of particles in each bin.}
\par\vspace{0pt}
\centering
\label{fig:EkinProd_vs_EkinDet}
\end{figure}

Beside the secondary particles yield (Section~\ref{sec:flux}) and emission profile (Section~\ref{sec:profiles}), the kinetic energy distribution of secondary particles is a crucial information to be exploited for range monitoring purposes. Charged secondary particles cross several centimeters of patient's tissue before exiting the body, losing kinetic energy and undergoing multiple scattering (MS). Therefore modeling and quantifying these effects is one of the challenges of ion beam therapy monitoring based on charged secondaries detection.

In this study, the detected kinetic energy of the secondary particles measured after they exit the PMMA target ($\rm E_{kin}^{Det}$) is reported. The detected kinetic energy ($\rm E_{kin}^{Det}$) can be related to the proton kinetic energy at production ($\rm E^{Prod}_{kin}$), considering the energy loss in the PMMA, as shown in Fig.\,\ref{fig:EkinProd_vs_EkinDet}, obtained from the high statistics MC simulation. $\rm 3.0 \times 10^{9}$ protons were produced in the PMMA target uniformly in z direction and isocentrically in the tranversal plane, with the FWHM=1\,cm and in energy range 10-250\,MeV. The uncertainty on the transformation from $\rm E^{Prod}_{kin}$ to $\rm E_{kin}^{Det}$ results mainly from the beam spot size (Tab.~\ref{tab:beams}) and as a consequence from the distance in the PMMA material that particles have to go through to exit the target.

In order to use secondary protons for monitoring purposes, the crossing of some centimeters of patient’s tissue has to be considered and therefore the range $\rm E_{kin}^{Prod} >$~60\,MeV (relative to $\sim$2.5\,cm) of the detected kinetic energy distribution is the most interesting for the above-mentioned application~\cite{Agodi2012b}.

\begin{figure}[h]
    \subfigure[\car ~at 120\,MeV/u at 90\degree] {\includegraphics[trim=0.cm 0.cm 0.cm 1.5cm, clip, width=0.3\textwidth]{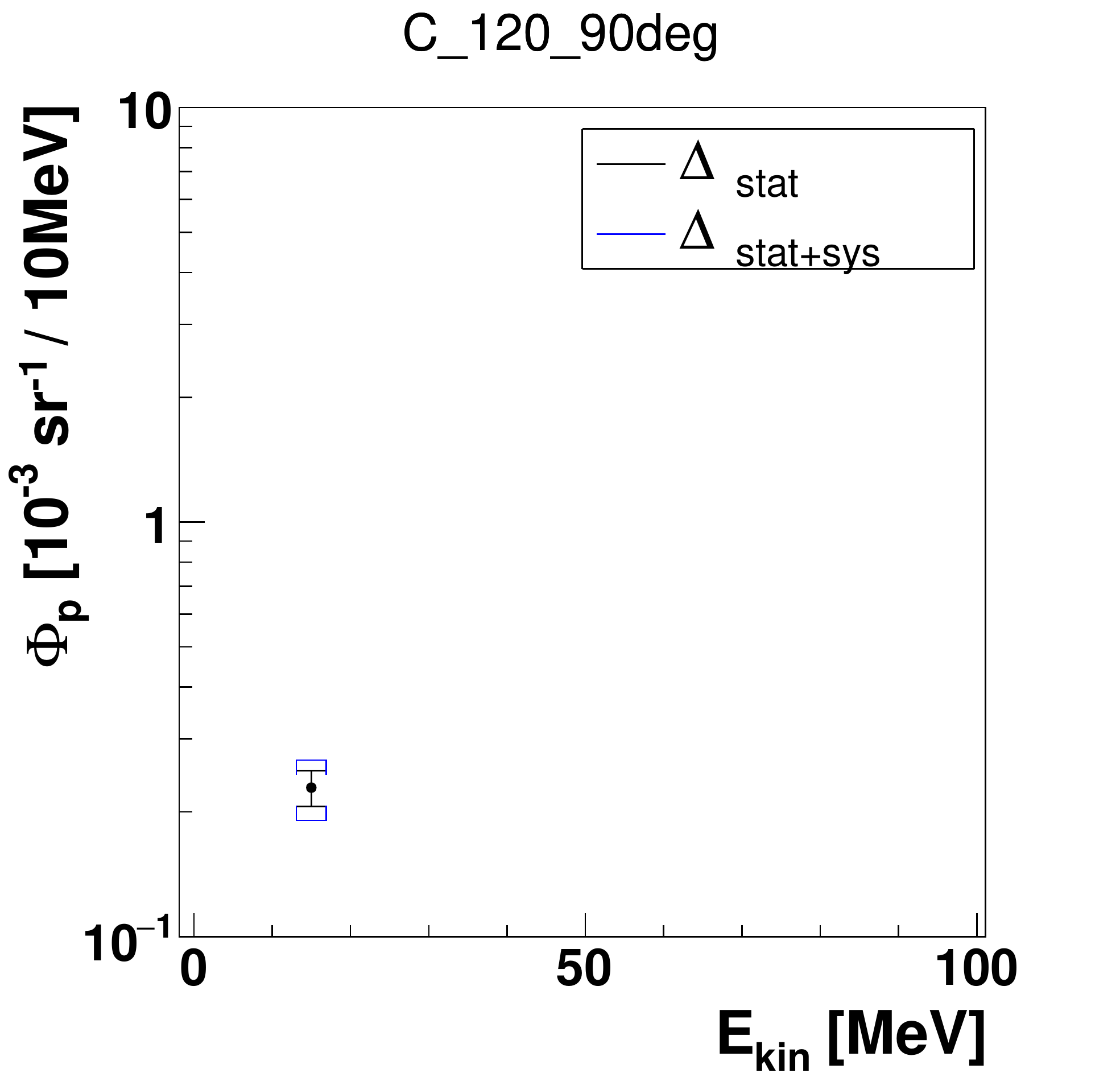}}
    \subfigure[\car ~at 160\,MeV/u at 90\degree] {\includegraphics[trim=0.cm 0.cm 0.cm 1.5cm, clip, width=0.3\textwidth]{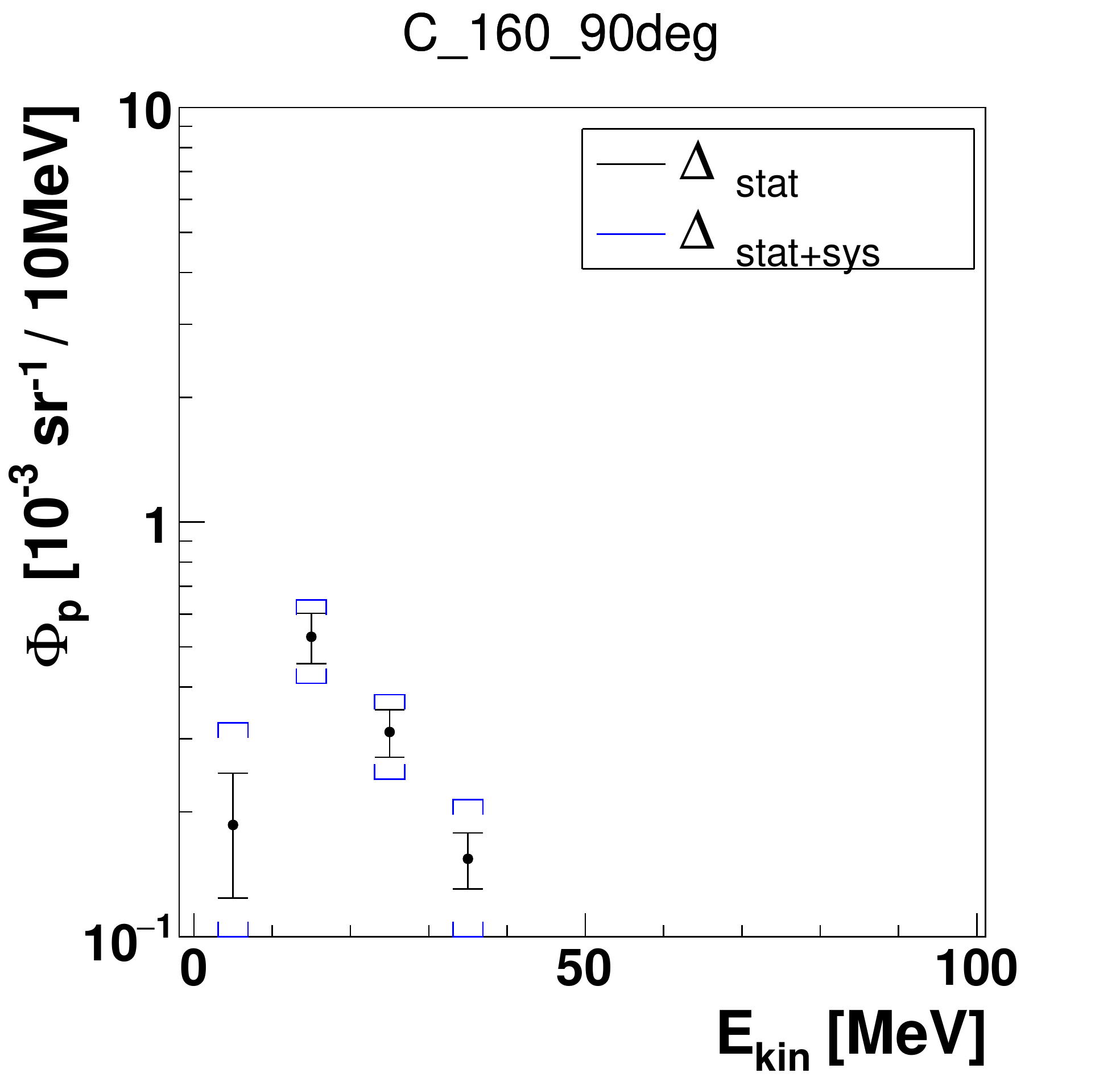}}
    \subfigure[\car ~at 180\,MeV/u at 90\degree] {\includegraphics[trim=0.cm 0.cm 0.cm 1.5cm, clip, width=0.3\textwidth]{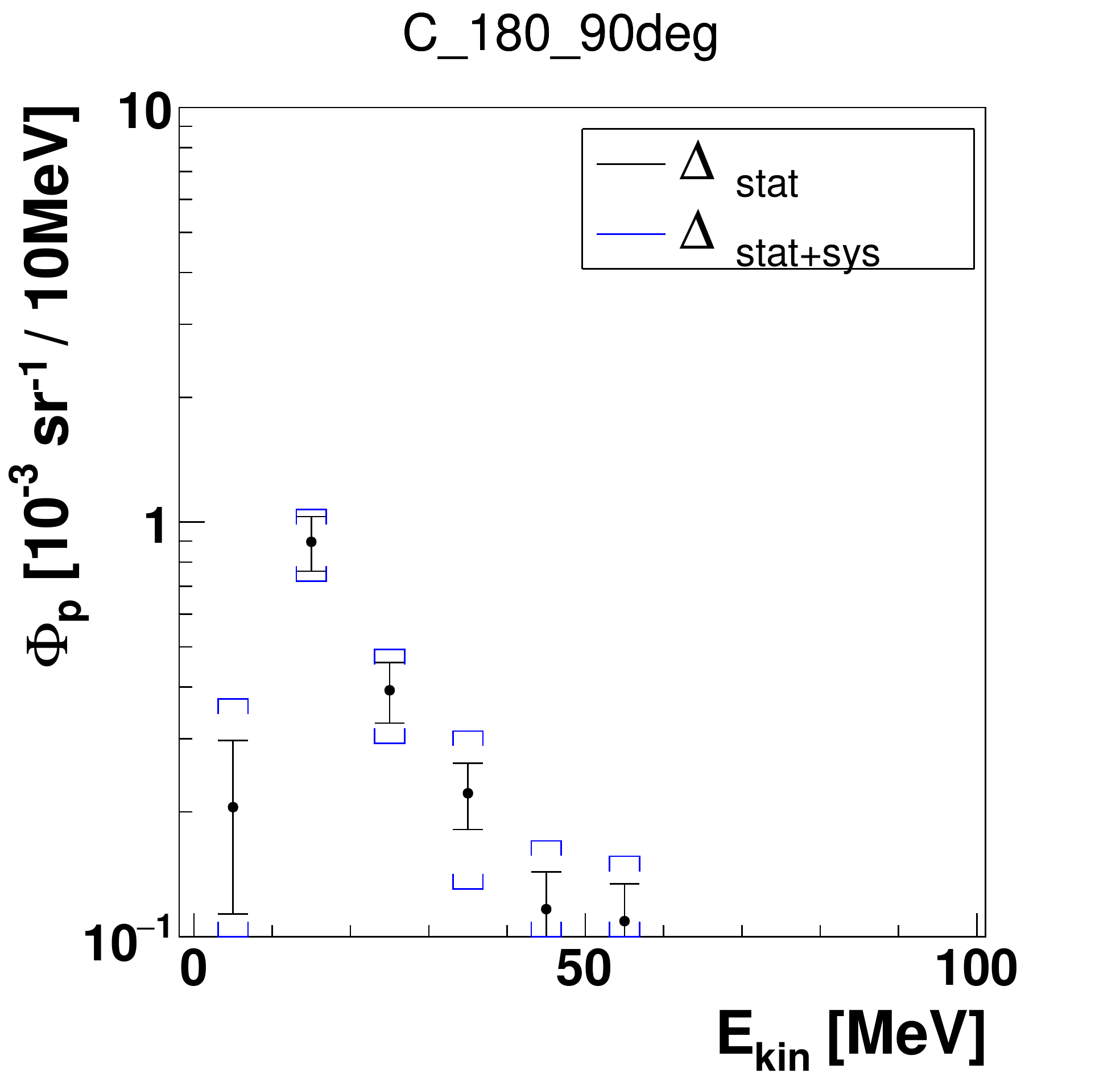}}
\newline
    \subfigure[\car ~at 220\,MeV/u at 90\degree] {\includegraphics[trim=0.cm 0.cm 0.cm 1.5cm, clip, width=0.3\textwidth]{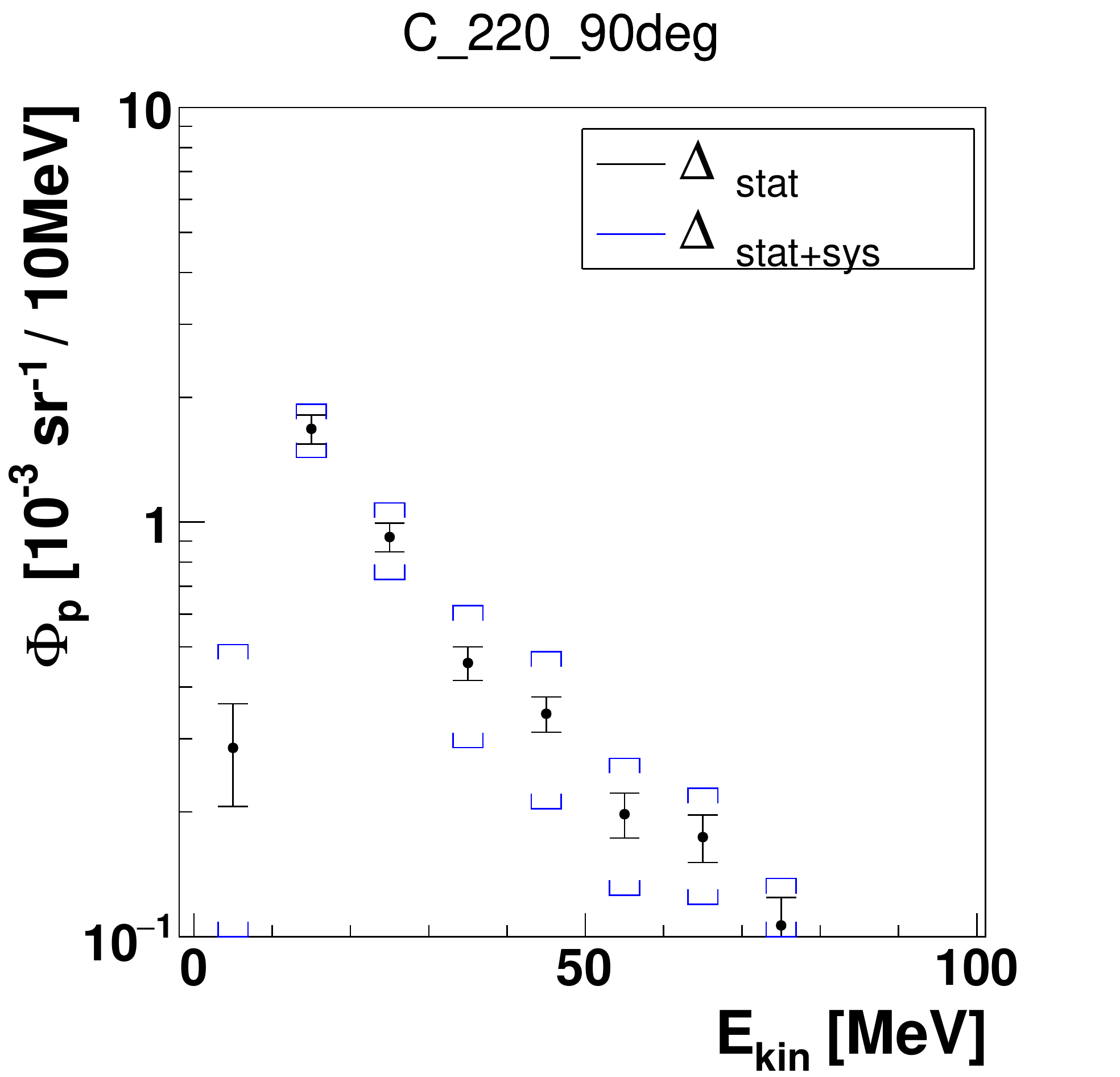}}
    \subfigure[\he ~at 125\,MeV/u at 90\degree] {\includegraphics[trim=0.cm 0.cm 0.cm 1.5cm, clip, width=0.3\textwidth]{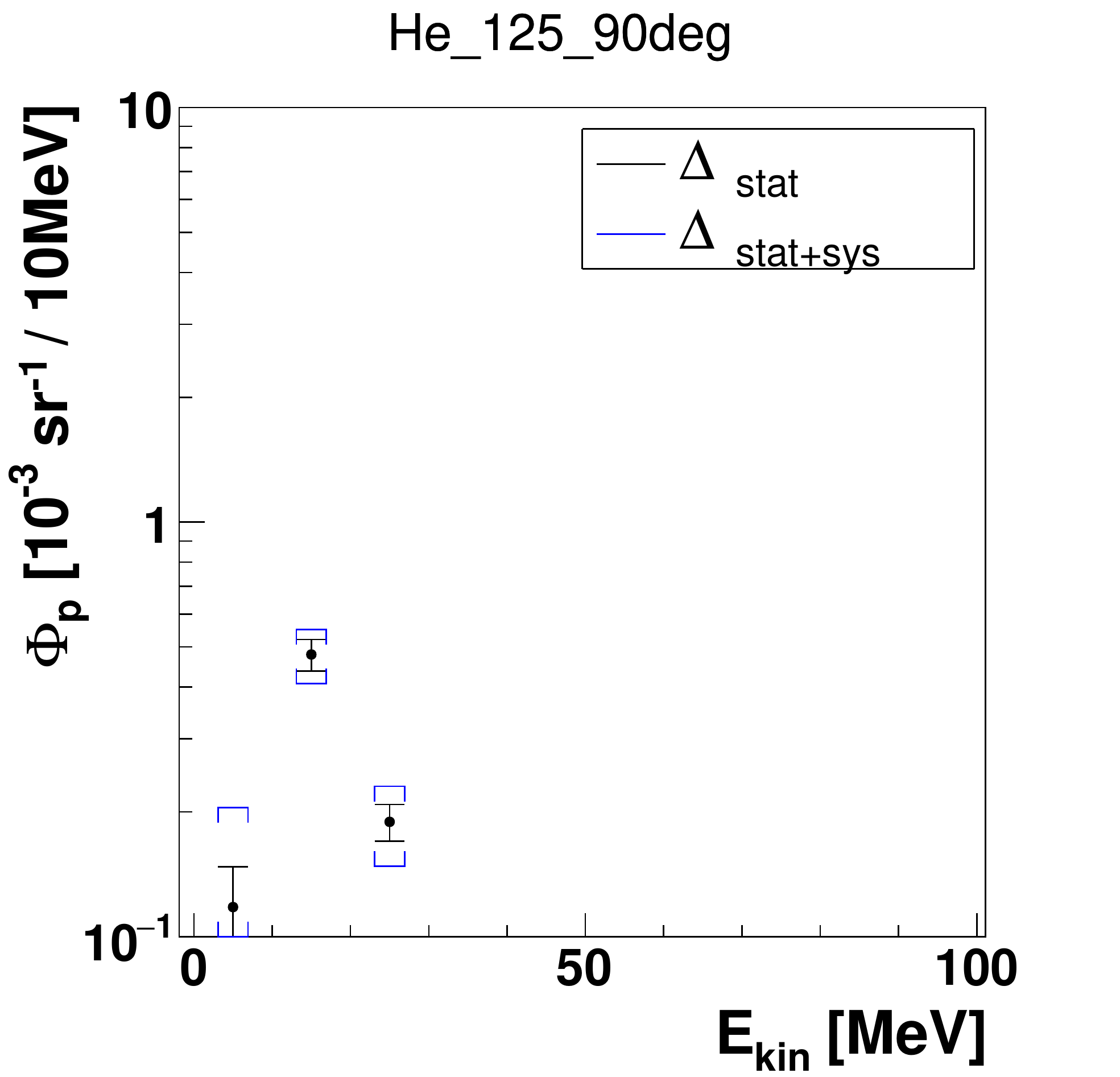}}
    \subfigure[\he ~at 145\,MeV/u at 90\degree] {\includegraphics[trim=0.cm 0.cm 0.cm 1.5cm, clip, width=0.3\textwidth]{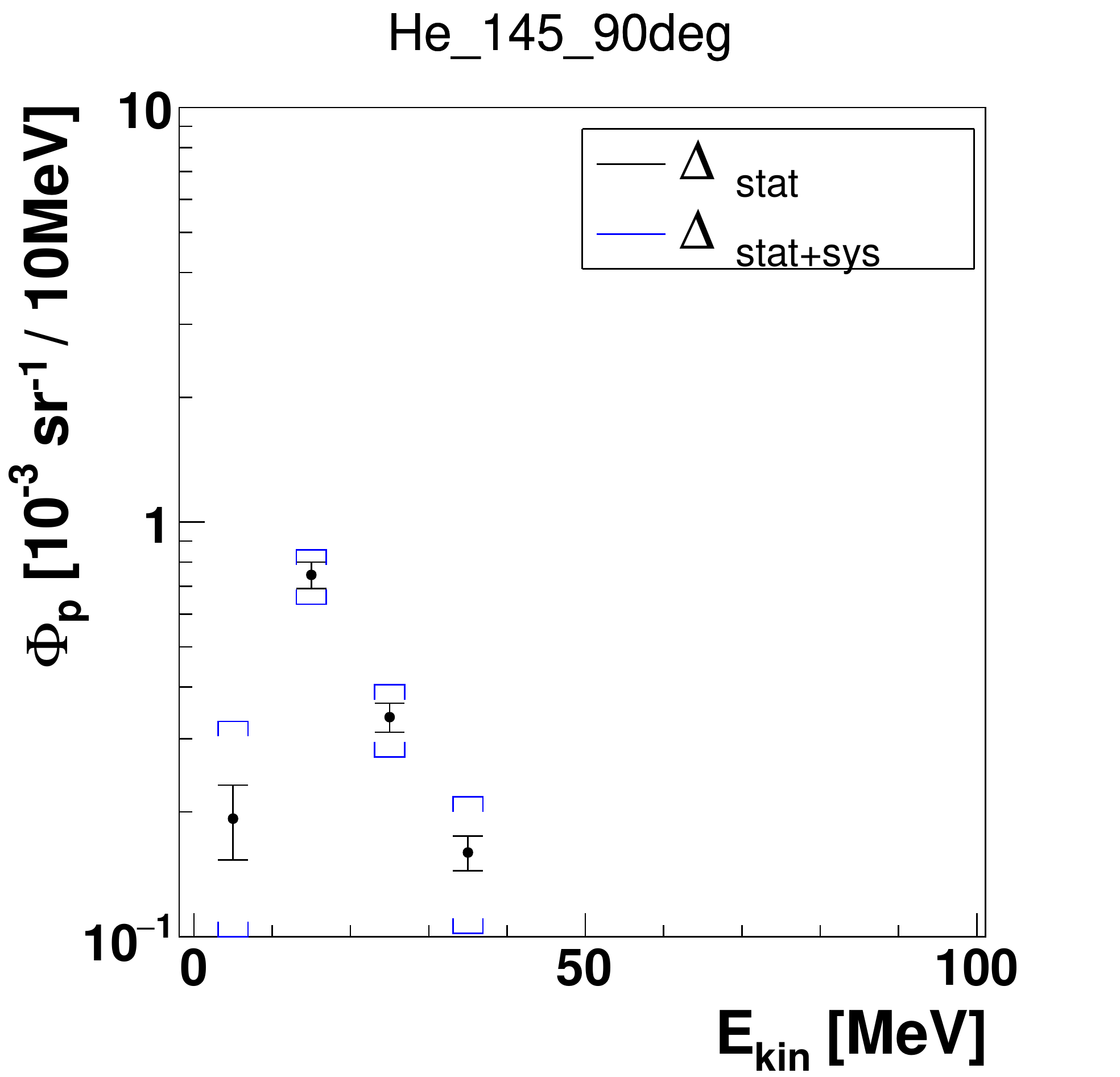}}
    \subfigure[\he ~at 102\,MeV/u at 60\degree] {\includegraphics[trim=0.1cm 0.cm 0.cm 1.5cm, clip, width=0.3\textwidth]{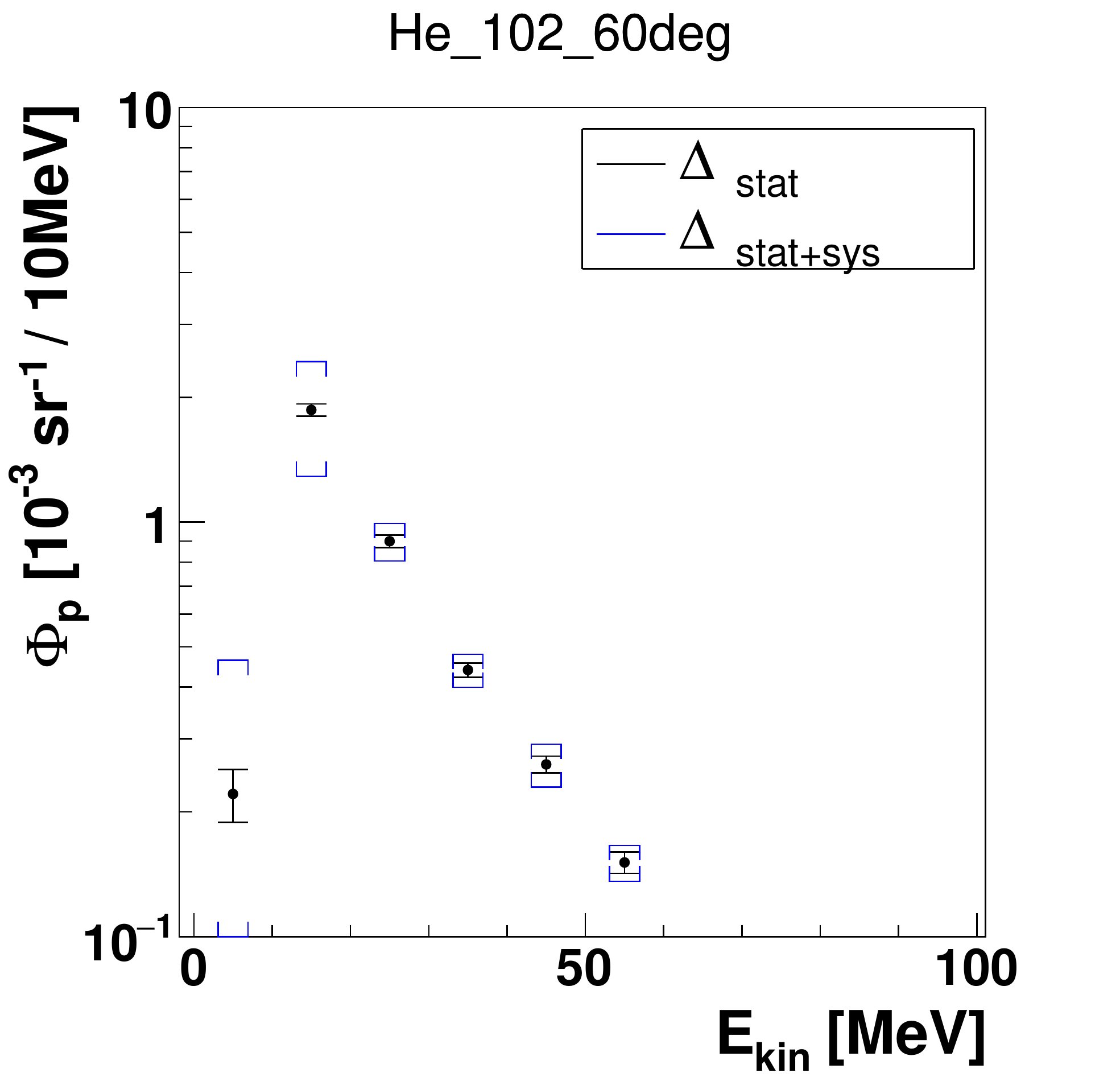}}
    \subfigure[\he ~at 125\,MeV/u at 60\degree] {\includegraphics[trim=0.cm 0.cm 0.cm 1.5cm, clip, width=0.3\textwidth]{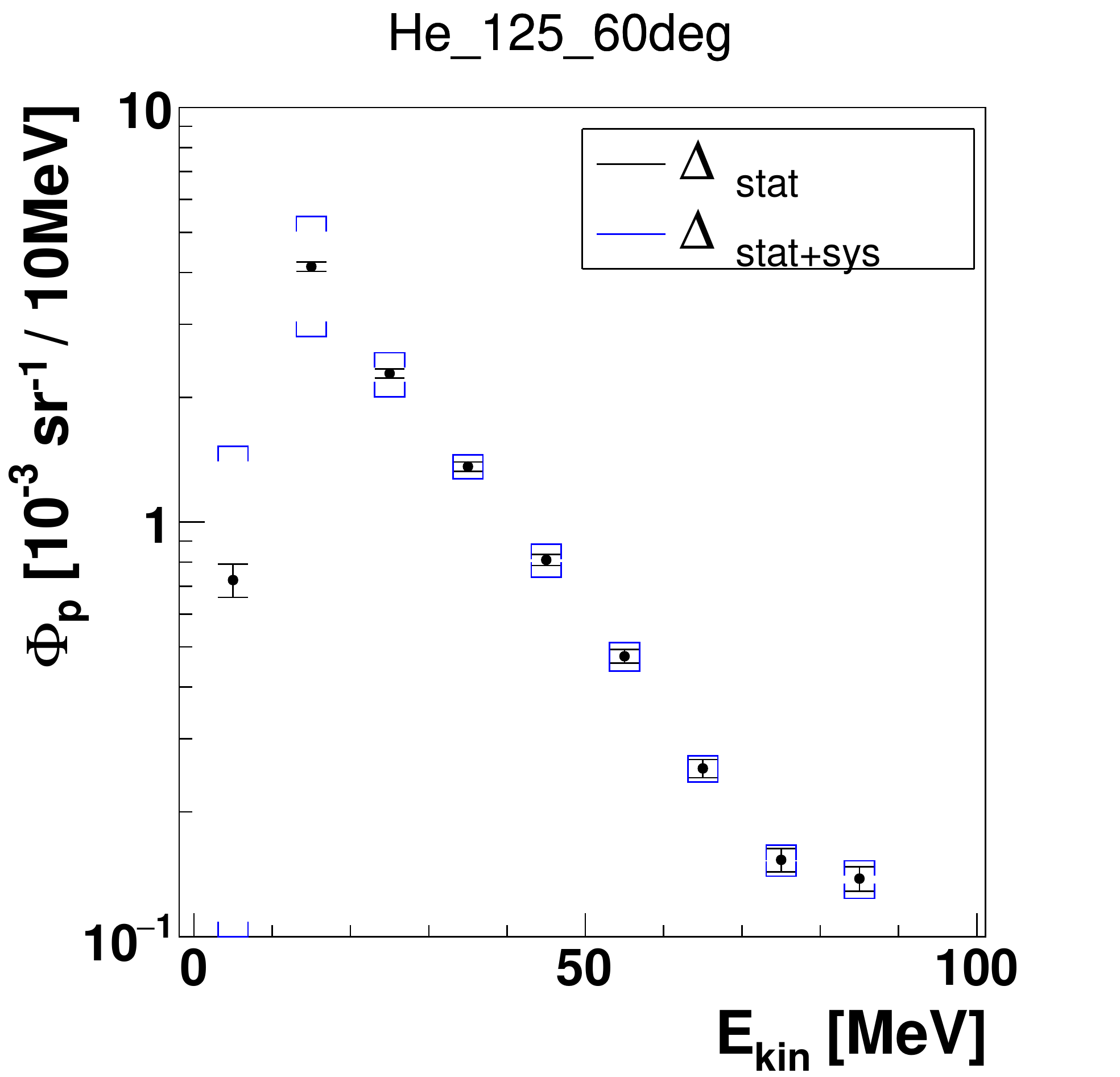}}
    \subfigure[\he ~at 145\,MeV/u at 60\degree] {\includegraphics[trim=0.cm 0.cm 0.cm 1.5cm, clip, width=0.3\textwidth]{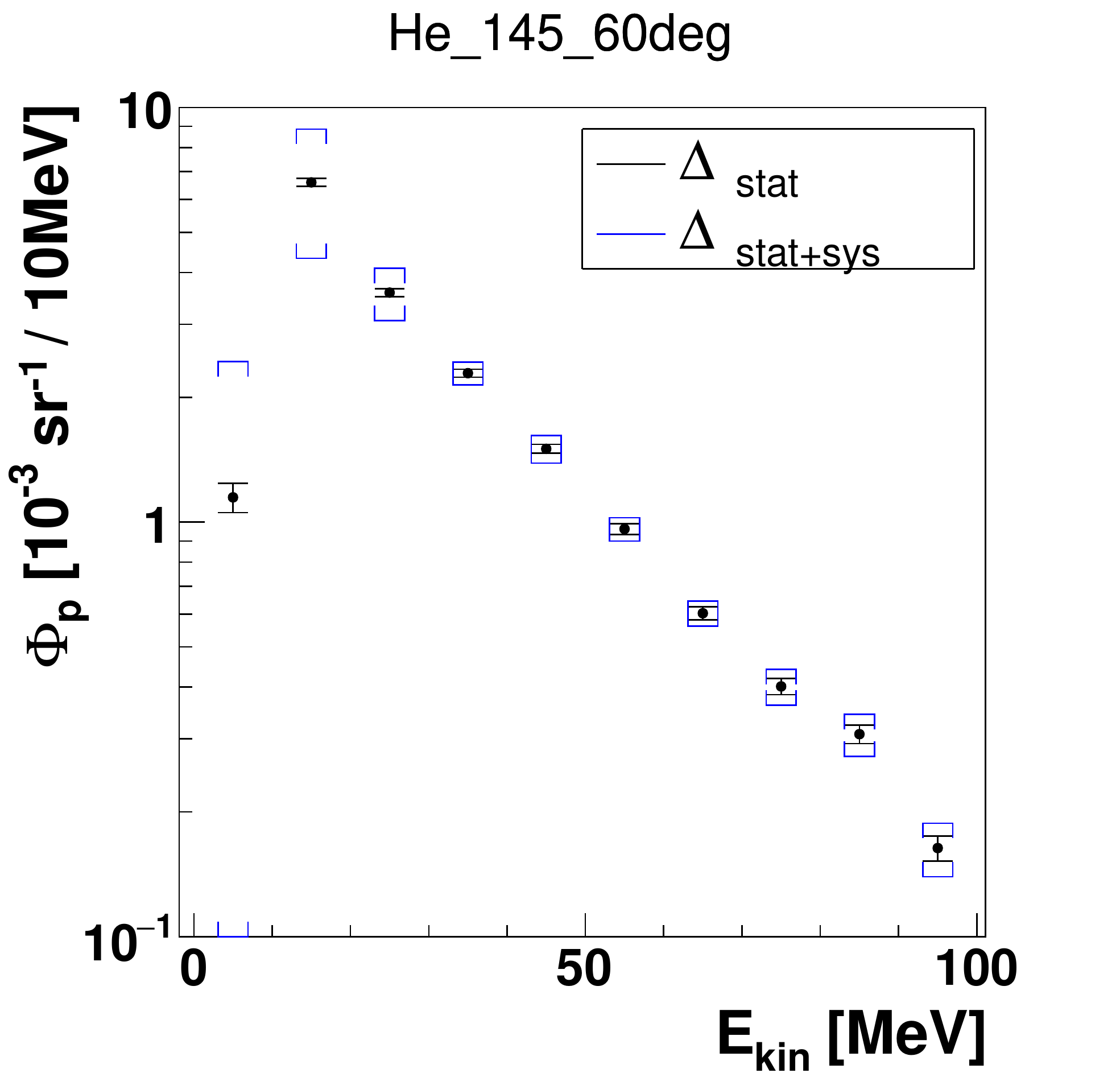}}
    \caption{Yield of charged secondary protons produced by \car ~and \he ~beams as a function of the detected kinetic energy $E_{\rm kin}$. Statistical ($\rm \Delta_{stat}$) and statistical+systematic ($\rm \Delta_{stat+sys}$) uncertainties are reported.}
\label{fig:ekinSpectra}
\end{figure}

Fig.~\ref{fig:ekinSpectra} shows the measured yields of secondary protons for \car ~ion beam and \he ~ion beam as a function of their detected kinetic energy $\rm E_{kin}^{Det}$, obtained using the TOF measurement performed using LTS and LYSO crystals signals. For each primary beam energy, the yield integrated over all kinetic energies of secondary protons in Fig.~\ref{fig:ekinSpectra} is equal to the total yield reported in Tab.~\ref{tab:fluxesP}. The number of secondary particles produced in the target increases and their energy spectrum widens with the energy of the primary beam, i.e. with its range. Comparing beams having a similar range (Fig.~\ref{fig:ekinSpectra}\,d,e), the yield is higher and the energy spectrum of charged secondary protons produced by the \car ~ion beam at 220\,MeV/u (Fig.~\ref{fig:ekinSpectra}\,d) extends to higher energies than the one produced by \he ~beam at 125\,MeV/u (Fig.~\ref{fig:ekinSpectra}\,e), as the secondary particles are produced essentially in projectile fragmentation. The secondary proton yield induced by \he ~beam and measured at 60\degree ~ with respect to the primary beam direction (Fig.~\ref{fig:ekinSpectra}\,g,h,i) is one order of magnitude higher than the one measured at 90\degree ~(Fig.~\ref{fig:ekinSpectra}\,e,f).

\section{Emission profiles}
\label{sec:profiles}

\begin{figure}
    \centering{}
    \subfigure[\car ~beam, 90\degree configuration.] {\includegraphics[trim=0.cm 0.cm 0.cm 0.cm, clip,  width=0.32\textwidth]{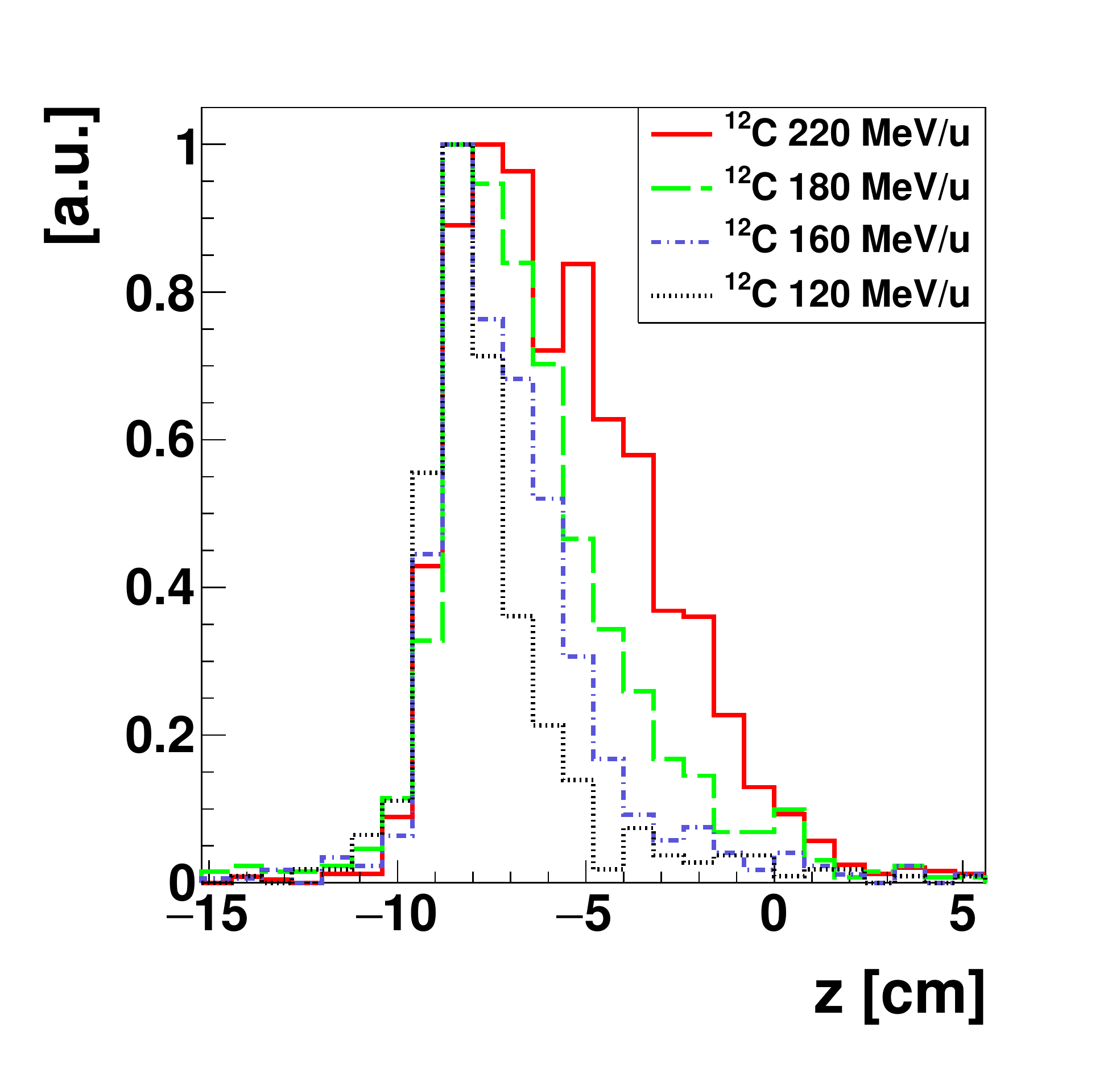}}
    \subfigure[\he ~beam, 90\degree configuration.] {\includegraphics[trim=0.cm 0.cm 0.cm 0.cm, clip, width=0.32\textwidth]{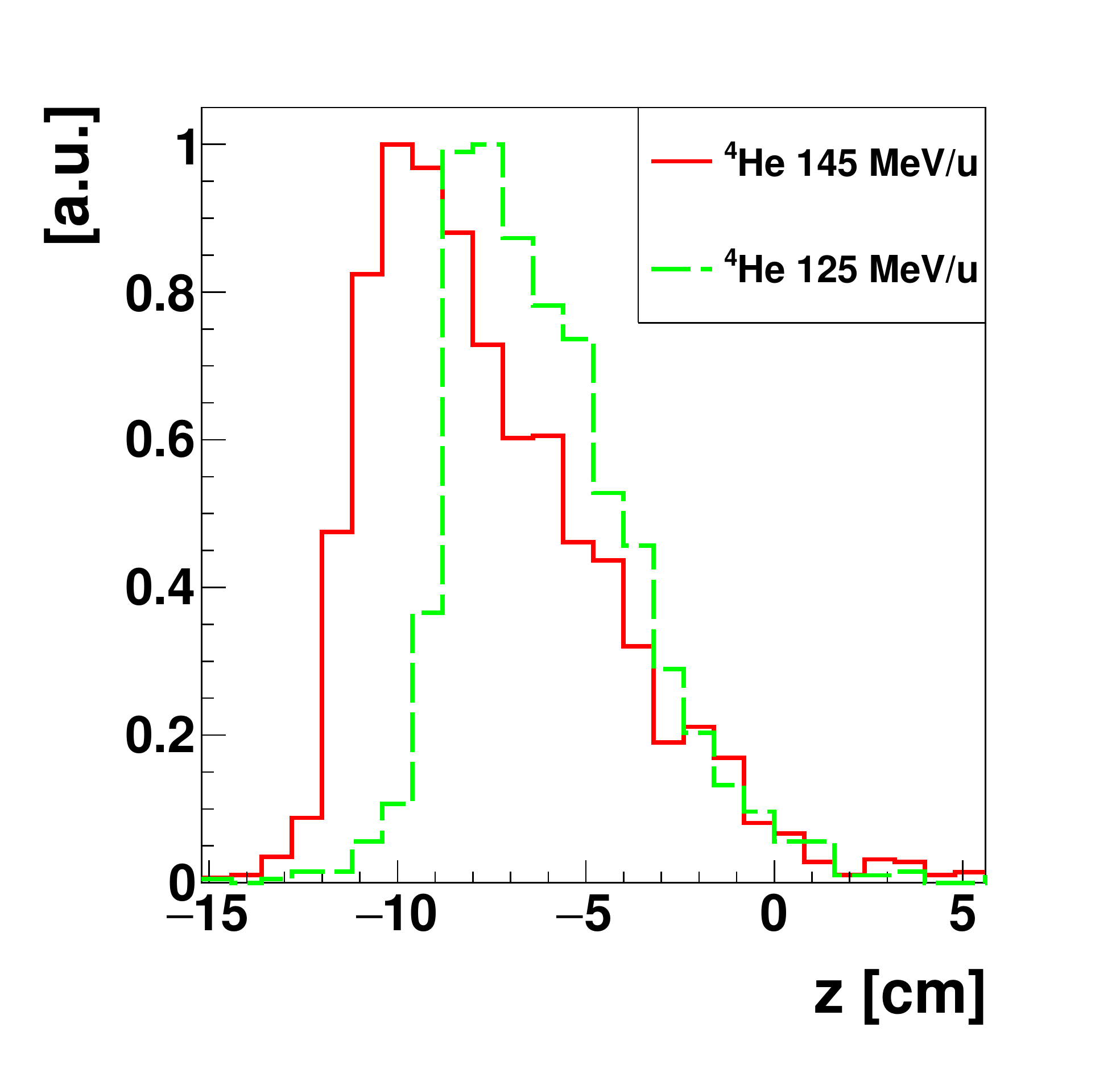}}
    \subfigure[\he ~beam, 60\degree configuration.] {\includegraphics[trim=0.cm 0.cm 0.cm 0.cm, clip, width=0.32\textwidth]{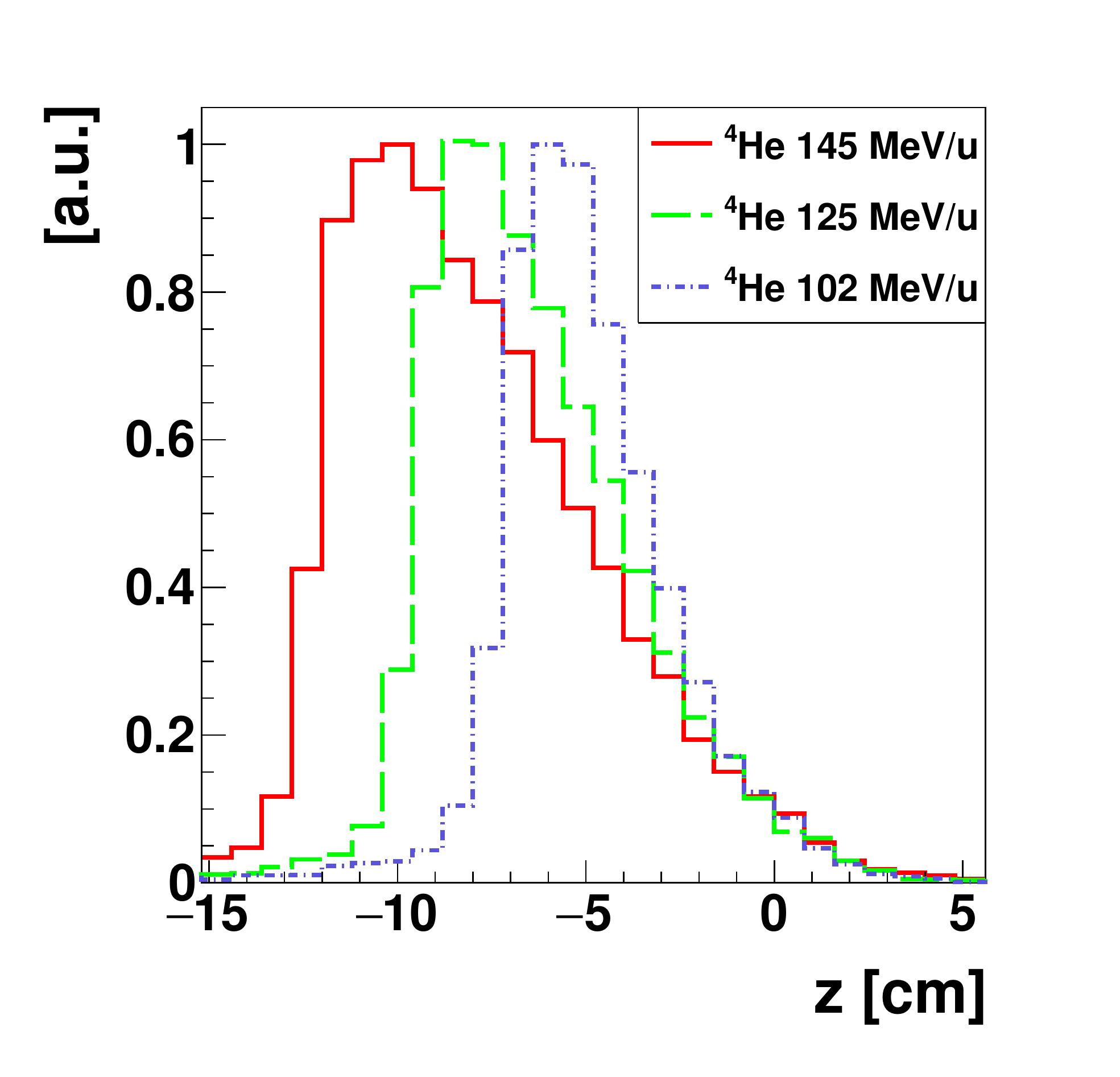}}
        \subfigure[\he ~beam at 125\,MeV/u, 90\degree configuration.] {\includegraphics[trim=0.cm 0.cm 0.cm 0.cm, clip, width=0.48\textwidth]{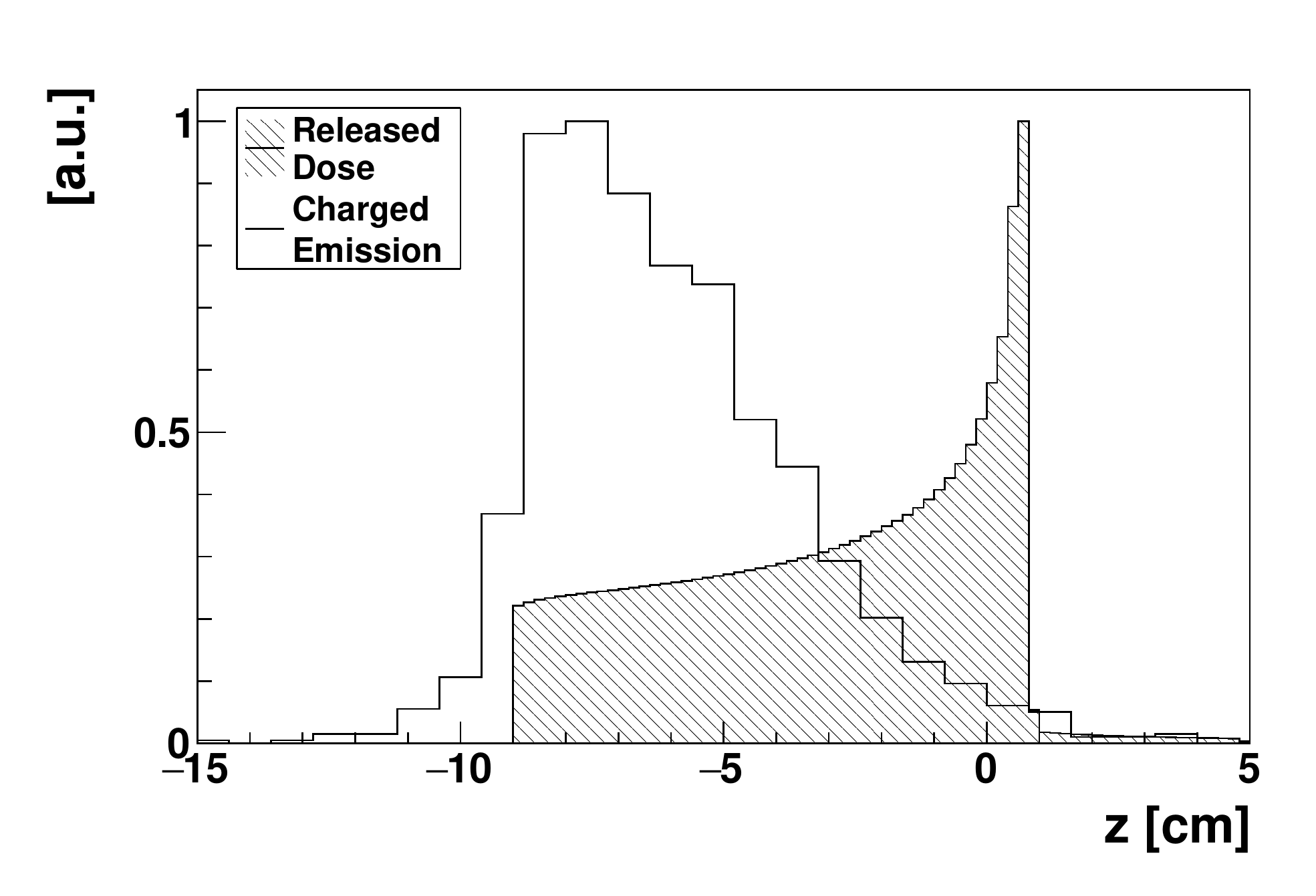}}
    \subfigure[\he ~beam at 125\,MeV/u, 90\degree configuration.] {\includegraphics[trim=0.cm 0.cm 0.cm 0.cm, clip, width=0.48\textwidth]{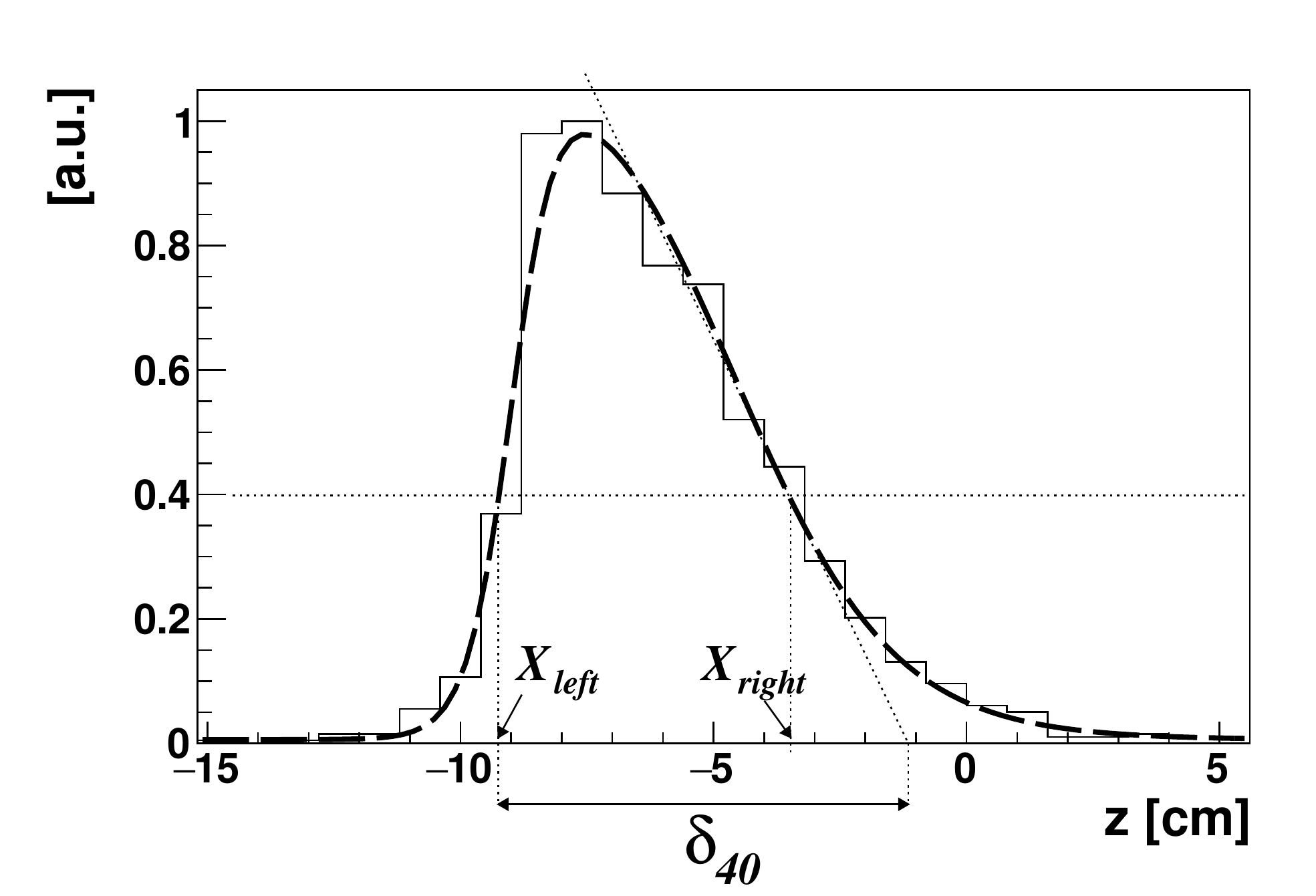}}
    \caption{Longitudinal profile of charged secondary fragments reconstructed inside the PMMA target. At Fig.\,(a,d,e) the beam entrance face is at -9.0\,cm, whereas at Fig.\,(b)\,and\,(c) the beam entrance face is at -11.7, -9.0, -6.7\,cm for \he ~beam at 145, 125 and 102\,MeV/u, respectively.
Fig.\,(d) illustrates the z-profile (Charged Emission - solid line) for \he ~beam at 125\,MeV/u detected at 90\degree ~(middle distribution from the Fig.\,b) and the corresponding dose released inside the target (Released Dose - hatched area). The parameters of the emission profile shown in Fig.\,(e) were estimated based on 40\% threshold (horizontal dotted line) as introduced in~\cite{Piersanti2014}.}
  \label{fig:profiles}
\end{figure}

Using the DCH information, the charged secondary particles were back-tracked to the PMMA target. The longitudinal emission profile of charged secondaries produced by the primary therapeutic beam (z-profile; Fig.~\ref{fig:profiles}) was built by considering all the reconstructed tracks. The correlation between BP position of a \car ~ion beam at 220\,MeV/u and charged secondary emission profile has already been shown before~\cite{Agodi2012b,Piersanti2014}. The emission spectra were investigated for \car ~ion and \he ~ion beams at all the energies. As an example Fig.~\ref{fig:profiles}d shows the dose released by \he ~ion beam at 125\,MeV/u overlapped with the reconstructed z-profile.

For each ion beam energy, the emission profile of charged secondaries was reconstructed and a fit implemented using a chisquare minimization was performed using a Double Fermi Dirac (DFD) function (Fig.~\ref{fig:profiles}e) as introduced in~\cite{Piersanti2014}:

\begin{equation} 
\rm f(x)=p_{0}\frac{1}{1+exp(\frac{z-p_{1}}{p_{2}})}\frac{1}{1+exp(-\frac{z-p_{3}}{p_{4}})}+p_{5}. 
\label{eq:dfd}
\end{equation}
The fit parameters $\rm p_{3}$ and $\rm p_{1}$ are respectively related to the position of the rising and falling edge of the distribution, while $\rm p_{4}$ and $\rm p_{2}$ describe the rising and falling slopes of the function, whereas $\rm p_{5}$ models a flat background contribution.
The parameters of the distribution characterizing the emission shape are shown in the Fig.~\ref{fig:profiles}e, extracted and listed in Tab.~\ref{tab:eProfiles_xleft}~and~\ref{tab:eProfiles_delta40} for different ion beams and beam energies. The parameters $\rm X_{left}$, $\rm X_{right}$ and $\rm \delta_{40}$ were calculated at 40\% of the maximum of DFD function~\cite[Fig.~\ref{fig:profiles}e: horizontal dotted line]{Piersanti2014}. The $\rm X_{left}$ parameter corresponds to the rising edge of the emission shape and indicates the PMMA entrance face position ($\rm EF_{PMMA}$) and the $\rm \delta_{40}$ parameter is correlated to the range (R) of the primary beam. The uncertainty on $\rm X_{left}$ and $\rm \delta_{40}$ parameter is related to the sample statistics used to obtain the emission shape (cf.~Section\,\ref{sec:dataSelectionAndPid}). The uncertainty on $\rm EF_{PMMA}$ and $\rm R$ are negligible.

The charged secondary emission shape varies with the primary ion beam energy for both \car ~and \he ~(Fig.~\ref{fig:profiles}).
The \car ~beam, at each of the investigated energies, entered the 10\,cm long PMMA target at the same position as it is indicated by the rising edge of the emission profile ($\rm X_{left}$; Fig.~\ref{fig:profiles}a; Tab.~\ref{tab:eProfiles_xleft}). Decreasing the energy of the \car ~ion beam, the emission profile becomes shorter and the slope of its falling edge becomes steeper, as the production of the secondaries decreases with the range of the primary ion beam (see R and $\rm \delta_{40}$ parameter in Tab.~\ref{tab:eProfiles_delta40}). 
Differently from the \car ~beam, for each energy of \he ~beam the length of the PMMA target was adapted in such a way that the BP position was before the distal end of the target (see Tab.~\ref{tab:beams}). The beam entrance face was at different positions as indicated by the rising edge of the emission profile (Fig.~\ref{fig:profiles}\,b,c) and the $\rm X_{left}$ parameter value of the emission shape (Tab.~\ref{tab:eProfiles_xleft}).

\begin{table}
\centering
\caption{Emission shape parameter (X$_{\rm \textbf{left}}$) extracted from the fit of the emission shape calculated with a Double Fermi Dirac function and related to the expected entrance face (EF$_{\rm PMMA}$) of the PMMA target. \\}
\label{tab:eProfiles_xleft}
\resizebox{\textwidth}{!}{
\begin{tabular}{ccc | ccc | c }
$\bm{ \theta}$ & \textbf{Ion} & \textbf{Energy} & \textbf{EF$_{\rm PMMA}$} & \textbf{X$_{\rm \textbf{left}}$} & \textbf{EF$_{\rm PMMA}$-X$_{\rm \textbf{left}}$} & \textbf{(EF$_{\rm PMMA}$-X$\rm _{\textbf{left}})_{calib}$} \\ 
 & (MeV/u) &  & (cm) & (cm) & (cm) & (cm) \\ \hline
\multirow{4}{*}{90\degree} & \multirow{4}{*}{\car}  & 120 & \multirow{4}{*}{-9.0±0.1} & -9.4±0.2 & 0.4 & 0.1±0.2 \\
 & & 160 & & -9.3±0.1 & 0.3 & 0.0±0.1  \\
 & & 180 & & -9.1±0.1 & 0.1 & -0.1±0.1 \\
 & & 220 & & -9.3±0.1 & 0.3 & 0.0±0.1 \\
  & & & & & calib=0.3 &  \\ \hline
\multirow{2}{*}{90\degree} & \multirow{2}{*}{\he} & 125 & -9.0±0.1 & -9.2±0.1 & 0.2 & 0.1±0.1 \\
 & & 145 & -11.7±0.1 & -11.7±0.1 & 0.0 & -0.1±0.1 \\
   & & & & & calib=0.1 &  \\ \hline
\multirow{3}{*}{60\degree}& \multirow{3}{*}{\he} & 102 & -6.7±0.1 & -7.5±0.1 & 0.8 & 0.0±0.1 \\
 & & 125 & -9.0±0.1 & -9.8±0.1 & 0.8 & 0.0±0.1  \\
 & & 145 & -11.7±0.1 & -12.5±0.1 & 0.8 & 0.0±0.1 \\ 
    & & & & & calib=0.8 &  \\ \hline
\end{tabular}
}
\vspace{1.4cm}
\centering
\caption{Emission shape parameter ($\rm \delta_{\textbf{40}}$) extracted from the fit of the emission shape calculated with a Double Fermi Dirac function and related to the primary ion beam range (R) in the PMMA target. \\}
\label{tab:eProfiles_delta40}
\resizebox{0.85\textwidth}{!}{
\begin{tabular}{ccc | ccc | c }
$\bm{ \theta}$ & \textbf{Ion} & \textbf{Energy} & \textbf{R} & $\rm \delta_{\textbf{40}}$ & \textbf{R-$\rm \delta_{\textbf{40}}$ } & \textbf{(R-$\rm \delta_{\textbf{40}})_{calib}$} \\ 
 & (MeV/u) &  & (cm) & (cm) & (cm) & (cm) \\ \hline
\multirow{4}{*}{90\degree} & \multirow{4}{*}{\car} & 120 & 2.9 & 3.8±0.5 & 0.9 & 0.0±0.5 \\
 & & 160 &  4.8   & 5.6±0.8 & 0.8 & -0.1±0.8 \\
 & & 180 &  6.0 & 6.8±0.6 & 0.8 & -0.1±0.6 \\
 & & 220 &  8.3 & 9.3±0.4 & 1.0 & 0.1±0.4 \\
     & & & & & calib=0.9 &  \\ \hline
\multirow{2}{*}{90\degree} & \multirow{2}{*}{\he} & 125 & 9.7 & 8.1±0.5 & -1.6 & - \\
 & & 145 & 12.5 & 10.5±0.5 & -2.0 & - \\ \hline
\multirow{3}{*}{60} & \multirow{3}{*}{\he} & 102 & 6.7 & 6.8±0.3 & 0.1 & - \\ 
 & & 125 & 9.7 & 9.2±0.2 & -0.5 & - \\
 & & 145 & 12.5 & 11.8±0.2 & -0.7 & - \\ \hline
\end{tabular}
}
\end{table}

In order to prove the feasibility of range monitoring with charged secondary profiles, the emission shape parameters were extracted. The difference between the expected and measured PMMA entrance face position $\rm (EF_{PMMA}-X_{left})$ as well as the difference between the expected and measured BP position $\rm (R-\delta_{40})$ are listed in Tab.~\ref{tab:eProfiles_xleft}~and~\ref{tab:eProfiles_delta40}, respectively. The reconstructed z-profile and corresponding parameters $\rm X_{left}$ and $\rm \delta_{40}$ vary depending on the ion species and angular configuration of the detector, as an effect of MS. For this reason a calibration ($_{\textit{calib}}$) must be considered and applied separately for \car ~and \he ~beams at 90\degree and 60\degree. A calibration offset was calculated as the average of $\rm (EF_{PMMA}-X_{left})$ and $\rm (R-\delta_{40})$ differences. The calibrated differences $\rm (EF_{PMMA}-X_{left})_{calib}$ and $\rm (R-\delta_{40})_{calib}$ are listed in the last column of Tab.~\ref{tab:eProfiles_xleft}~and~\ref{tab:eProfiles_delta40}. 

The calibrated difference between the expected and measured PMMA entrance face position $\rm (EF_{PMMA}-X_{left})_{calib}$ is within the uncertainty on the $\rm X_{left}$ parameter evaluation for all the emission profiles. The calibrated difference between the expected and measured BP position $\rm (R-\delta_{40})_{calib}$ is within the uncertainty on the $\rm \delta_{40}$ parameter evaluation for \car~emission profiles. For emission profiles obtained with \he ~beams the calibrated difference between the expected and measured BP position $\rm (R-\delta_{40})_{calib}$ exceeds the uncertainty on  the $\rm \delta_{40}$ parameter evaluation. Even if the relation between R and $\rm \delta_{40}$ is evident for \he ~beams, the selection of the parameter different than $\rm \delta_{40}$ or calibration of the $\rm (R-\delta_{40})$ relation as a function of energy might be necessary, when considering clinical application of range monitoring with charged secondary particles.

We confirm the feasibility of identifying patient mispositioning by estimating the $\rm X_{left}$ parameter for \car ~and \he ~beams. Furthermore, we confirm the feasibility of range monitoring by estimating $\rm \delta_{40}$ parameter by \car ~beams at different energies. The outcomes of the studies performed with \car ~ion beam confirm the findings reported by~\cite{Piersanti2014}. The collected data also indicate that charged secondary particles produced by \he ~beam could be used for range monitoring purposes in hadrontherapy. Further studies optimizing the emission profile parameters choice and their calibration as a function of the primary ion beam energy are needed. The accuracy of $\rm \delta_{40}$-based range monitoring depends mainly on MS of the fragments inside the patient, the statistics of collected sample and the detection angle with respect to the primary beam direction. The needed accuracy of a possible range monitoring device depends on the clinical treatment parameters that will be discussed in the next section.

\section{Discussion}
\label{sec:discussion}

This paper reports on the measurement and analysis of charged secondary particles produced by \he~and \car~ion beams impinging on a PMMA target. The measurements aimed to estimate secondary particle yields, energy spectra and emission shapes as a function of the primary beam energy in a range interesting for PT applications. Yields of charged secondary particles detected at 90\degree~and 60\degree~ with respect to the primary beam direction were obtained correcting for detection efficiency as a function of the kinetic energy, as well as the production position of the secondary particles. The secondary proton yield ranges from 0.5~to~17.5~$\rm \times 10^{-3}sr^{-1}$ per primary ion depending on the primary ion beam, its energy and setup configuration (90\degree~or~60\degree). We studied the statistical and systematic uncertainty on the yield and the total fractional uncertainty was estimated to be in the 12\%-22\% range. The energy spectra of charged secondary protons were plotted normalized to the number of primary ions and the uncertainty on the yield and particle TOF evaluation were indicated. The emission point of each detected secondary particle was reconstructed and the charged secondary emission profiles were built. These z-profiles were correlated to the expected range of the primary beam. The feasibility of range monitoring with a secondary particle tracking detector was confirmed for \car~ion beams at different energies in therapeutic range. The results obtained for the first time with \he ~beam suggest the feasibility of range monitoring based on charged secondary particle detection also for this beam.

In this study the homogeneous PMMA target was irradiated to characterize the production of secondary particles in therapeutic-like conditions. The lateral dimension of the target exceeded two times the largest beam FWHM used in the measurement ensuring that all the primary ions were stopped in the target. The maximal distance to be traveled by secondary particles to exit the PMMA is $\sim$3\,cm, which translates into a minimal proton kinetic energy at production $\rm E_{kin}^{Prod}=50\,MeV$.

In order to exploit dose monitoring techniques based on the detection of charged secondary particles in the clinical practice, the calibration of the detection device must be performed accounting for detector acceptance, detector position with respect to the primary beam and detector performance. In the clinical conditions one can consider as an example charged secondary detector of solid angle equal to 0.08 \cite[20x20\,cm field of view positioned 20\,cm from the patient]{Traini2016}. Below, an example of secondary proton yield calculation performed for a head and neck \car ~ion treatment plan follows. The example plan consists of 4500 raster points distributed over 40 energy slices. In total $2.7 \times 10^{9}$ particles were irradiated. The slice corresponding to primary \car ~beam at 220\,MeV/u consists of $\sim$100 raster points, each irradiated with $\sim6.0 \times 10^{6}$ particles assuming an equal distribution of the number of particles per raster point. Multiplying number of particles delivered per raster point by the detector solid angle (0.08) and secondary proton yield obtained for \car ~at 220\,MeV/u (Tab.~\ref{tab:fluxesP}) one expects to detect $\sim$2000 secondary protons. This secondary particle statistics corresponds to an uncertainty on emission profile detection of 4\,mm. This uncertainty could be substantially improved joining information from few neighboring raster points.

The translation of charged secondary monitoring technique to the clinical practice of PT requires systematic studies of the clinical scenarios considering the dependence of the resolution on $\rm X_{left}$ and $\rm \delta_{40}$ parameters estimation (i.e.,~charged secondary particles yield and energy spectra) on the size and location of the tumour as well as the prescribed treatment dose.

In the measurements performed in Heidelberg the data on charged secondary, prompt-photons, \betaplus production and forward fragmentation in PMMA target have been collected. Charged secondary yields, energy spectra and emission profiles produced with \he ~and ~\car ~beams is the subject of this publication, whereas the charged secondary production obtained with \oxi ~beam is the subject of further analysis. Design of the new tracking device for range monitoring based on the detection of charged secondary protons is proposed in ~\cite{Traini2016}. Prompt photon yields produced with \he, \car ~and \oxi ~beams have been reported in~\cite{Mattei2016}. 

\section{Conclusions}
\label{sec:conclusions}

The charged secondary yields, energy spectra and emission profiles produced by \car ~and \he ~ ion beams were studied at 90\degree and 60\degree. The obtained results confirm feasibility of ion beam therapy range monitoring using \car ~ion beam and suggest feasibility of range monitoring with \he ~beam. The simulation studies considering patient treatment plans and patient geometry are needed for the translation of the range monitoring technique based on the charge secondary detection to the clinic. 

\section*{Acknowledgements}
\label{sec:acknowledgements}

We would like to thank sincerely Marco Magi (SBAI Department) for his valuable effort in the construction of several mechanical supports of the experimental setup. This work has been partly supported by the "Museo storico della Fisica e Centro di studi e ricerche Enrico Fermi". The access to the test beam at the Heidelberg Ion- beam Therapy center has been granted by the ULICE European program. We are indebted to Prof.~Dr.~Thomas Haberer and Dr.~Stephan Brons for having encouraged this measurement, made possible thanks to their support and to the help of the whole HIT staff.

\section*{References}
\label{sec:references}

\end{document}